\documentclass[letterpaper, aps, prd, twocolumn, superscriptaddress, showpacs, nofootinbib,floatfix]{revtex4}
\pdfoutput=1

\usepackage{graphicx}
\usepackage{bm}
\usepackage{amssymb}
\usepackage{amsmath}
\usepackage{amsfonts}
\usepackage{dcolumn}
\usepackage{natbib}
\usepackage[normalem]{ulem} 
\usepackage[dvipsnames]{xcolor}
\usepackage{calc}
\usepackage{longtable}

\LTcapwidth=\textwidth

\newcommand{\z}{\phantom{0}}
\newcommand{\zz}{\phantom{00}}
\newcommand{\pz}{\phantom{.0}}
\newcommand{\p}{\phantom{.}}

\newcommand{\raiseentry}[1]{\smash{\raise 0.7 em \hbox{#1}}}

\newcommand{\lowentry}[1]{\smash{\lower 1.5 ex \hbox{#1}}}

\def\apj{Astrophys. J.}
\def\apjl{Astrophys. J. Lett.}
\def\apjs{Astrophys. J. Supp. Ser. }

\def\aap{Astron. Astrophys. }
\def\araa{Ann.\ Rev. Astron. Astroph. }

\def\physrep{Phys. Rep. }
\def\mnras{Mon. Not. Roy. Astron. Soc. }

\def\prl{Phys. Rev. Lett.}
\def\prd{Phys. Rev. D.}

\newenvironment{equationarray*}
{\arraycolsep 0.14 em
\begin{eqnarray*}}
{\end{eqnarray*}}

\begin{document}

\title{Measuring the Angular Momentum Distribution \\ in Core-Collapse
  Supernova Progenitors with Gravitational Waves}    

\author{Ernazar~Abdikamalov}
\email{abdik@tapir.caltech.edu}
\affiliation{TAPIR, MC 350-17, California Institute of Technology, 
1200 E California Blvd., Pasadena, CA 91125, USA}

\author{Sarah~Gossan}
\email{sarah.gossan@tapir.caltech.edu}
\affiliation{TAPIR, MC 350-17, California Institute of Technology, 
1200 E California Blvd., Pasadena, CA 91125, USA}

\author{Alexandra~M.~DeMaio}
\email{amd320@eden.rutgers.edu} 
\affiliation{Department of Physics and
  Astronomy, Rutgers, The State University of New Jersey, 136
  Frelinghuysen Road Piscataway, NJ 08854-8019, USA}
\affiliation{TAPIR, MC 350-17, California Institute of Technology,
  1200 E California Blvd., Pasadena, CA 91125, USA}

\author{Christian~D.~Ott}
\email{cott@tapir.caltech.edu}
\affiliation{TAPIR, MC 350-17, California Institute of Technology, 
1200 E California Blvd., Pasadena, CA 91125, USA}
\affiliation{Kavli IPMU (WPI), University of Tokyo, Kashiwa, Japan 277-8583}

\date{\today}


\begin{abstract}
  The late collapse, core bounce, and the early postbounce phase of
  rotating core collapse leads to a characteristic gravitational wave
  (GW) signal. The precise shape of the signal is governed by the
  interplay of gravity, rotation, nuclear equation of state (EOS), and
  electron capture during collapse. We explore the detailed dependence
  of the signal on total angular momentum and its distribution in the
  progenitor core by means of a large set of axisymmetric
  general-relativistic hydrodynamics core collapse simulations, in
  which we systematically vary the initial angular momentum
  distribution in the core. Our simulations include a microphysical
  finite-temperature EOS, an approximate electron capture treatment
  during collapse, and a neutrino leakage scheme for the postbounce
  evolution.  Our results show that the total angular momentum of the
  inner core at bounce and the inner core's ratio of rotational
  kinetic energy to gravitational energy $T/|W|$ are both robust
  parameters characterizing the GW signal. We find
  that the precise distribution of angular momentum is relevant only
  for very rapidly rotating cores with $T/|W|\gtrsim 8\%$ at bounce. We
  construct a numerical template bank from our baseline set of
  simulations, and carry out additional simulations to generate trial
  waveforms for injection into simulated advanced LIGO noise at a
  fiducial galactic distance of $10\,\mathrm{kpc}$.  Using matched
  filtering, we show that for an optimally-oriented source and
  Gaussian noise, advanced Advanced LIGO could measure the total
  angular momentum to within $\pm20\%$, for rapidly rotating
  cores. For most waveforms, the nearest known degree of precollapse
  differential rotation is correctly inferred by both our matched
  filtering analysis and an alternative Bayesian model selection
  approach. We test our results for robustness against systematic
  uncertainties by injecting waveforms from simulations utilizing a
  different EOS and and variations in the electron fraction in the
  inner core. The results of these tests show that these uncertainties
  significantly reduce the accuracy with which the total angular
  momentum and its precollapse distribution can be inferred from
  observations.
\end{abstract}


\maketitle


\section{Introduction}
\label{section:introduction}

Massive stars ($8 M_\odot \lesssim M \lesssim 130 M_\odot$, at zero
age main sequence [ZAMS]) undergo collapse at the end of their nuclear
burning lives once their electron-degenerate core exceeds its
effective Chandrasekhar mass. The inner core collapses subsonically
and, when its density exceeds that of nuclear matter, experiences core
bounce due to the stiffening of the nuclear equation of state (EOS). A
hydrodynamic shock forms at the interface of inner and
supersonically collapsing outer core. The shock quickly moves out, but
stalls within a few tens of milliseconds at a radius of
$100-200\,\mathrm{km}$, due to dissociation of infalling iron-group
nuclei and energy losses to neutrinos that stream away from the
semi-transparent region behind the shock \cite{bethe:90}. The shock
must be revived by some mechanism to drive an explosion and create the
spectacular display of a core-collapse supernova across the
electromagnetic spectrum.

For the vast majority of core-collapse supernovae with explosion
energies of $\sim 0.1-1\,\mathrm{B}$ ($1\, \mathrm{B}ethe =
10^{51}\,\mathrm{erg}$), the \emph{neutrino mechanism}
\cite{bethewilson:85,janka:12b,burrows:13a} is the favored mechanism
of shock revival. It relies on the deposition of a fraction of the
outgoing electron neutrino and electron antineutrino luminosity (with
a typical efficiency of order $10\%$) behind the stalled shock, but
also requires neutrino-driven convection and/or the standing accretion
shock instability (SASI; e.g., \cite{blondin:03}) to increase the
dwell time of accreted matter in the region behind the shock where net
energy absorption is possible. The neutrino mechanism fails in
spherical symmetry (1D, where convection and SASI are absent).  A
number of axisymmetric (2D) core-collapse supernova simulations with
detailed energy-dependent neutrino transport and microphysics now
report successful explosions \cite{bruenn:13,mueller:12a,mueller:12b},
but the first such 3D simulations are not yet conclusive
\cite{hanke:13,takiwaki:13a}. Other physics or effects such as
precollapse asphericities due to vigorous convective shell burning may
be needed for enabling robust explosions in 3D~\cite{couch:13d}.

There is, however, a class of highly-energetic core-collapse
supernovae with inferred explosion energies of up to $10\,\mathrm{B}$
that the neutrino mechanism alone seems too feeble to possibly
explain. This class includes relativistic Type Ic supernovae with
strongly Doppler-broadened spectral lines from compact
hydrogen/helium-poor progenitors (so-called Type Ic-bl supernovae;
e.g.~\cite{drout:11,smith:11}) and super-energetic Type II supernovae
from red supergiants (e.g., \cite{botticella:10,smith:12a}) and makes
up $1$-$2\%$ of all core-collapse supernovae \cite{smith:11}. All
supernovae associated with long gamma-ray bursts have been of Type
Ic-bl \cite{modjaz:11,hjorth:11}.

Such energetic events may require a central engine that can convert
the gravitational energy provided by collapse much more efficiently
into energy of the explosive outflow than neutrinos are capable of.
One possibility is the \emph{magnetorotational mechanism} in which a
millisecond-period protoneutron star with magnetar-strength magnetic
fields drives a jet-driven bipolar explosion
\cite{bisno:70,leblanc:70,wheeler:00,wheeler:02,burrows:07b}, which,
in some cases, might set the stage for a subsequent long gamma-ray
burst (e.g., \cite{wb:06,metzger:11}).

Current standard lore of stellar evolution theory and pulsar
birth-spin estimates state that most massive stars are rather slow
rotators at the end of their lives, having lost angular momentum to
stellar winds and not being strongly differentially rotating due to
angular-momentum redistribution by magnetic torques (e.g.,
\cite{heger:05,ott:06spin}).  Special conditions, such as chemically
homogeneous evolution at low metallicity \cite{woosley:06,yoon:06} or
binary interactions \cite{demink:13}, might be necessary to produce
the progenitors of hyper-energetic core-collapse supernovae and long
GRBs.

This may or may not be the case.  Current stellar evolutionary
calculations are still 1D and take into account rotation and angular
momentum loss and redistribution only approximately and in a
parameterized, non-self-consistent way. Pulsar birth-spin estimates,
which are based on magnetic-dipole radiation, could be off by large
factors if early spin-down occurred by direct conversion of spin
energy into magnetic field and/or kinetic energy of an explosive
outflow.  Keeping this in mind, it is not inconceivable that rotation
could play a significant role in many core-collapse supernovae.
Rotating core collapse naturally leads to differential rotation in the
outer protoneutron star and in the postshock region
\cite{ott:06spin,dessart:12a}. The free energy in differential
rotation\footnote{At fixed total angular momentum, uniform rotation is
  the lowest energy state. Any process capable of redistributing
  angular momentum will operate on differential rotation, driving a
  system towards uniform rotation.} could be tapped by the
magnetorotational instability (e.g.,
\cite{balbus:91,obergaulinger:09,akiyama:03}), which could either lead
to the growth of large-scale magnetic fields (via a dynamo; as argued
for by \cite{akiyama:03,burrows:07b}) or local dissipation (and
additional heating) by reconnection \cite{thompson:05}. Depending on
precollapse spin and magnetization, both possibilities could either
subdominantly assist the neutrino mechanism in reviving the shock, or
dominate the dynamics in a magnetorotational \cite{burrows:07b} or
magneto-viscous \cite{thompson:05,suzuki:08} explosion.

Gravitational waves (GWs) are the most direct and best probes of
rotation in stellar collapse and core-collapse supernovae. Rotation
naturally leads to a quadrupole (i.e.~oblate) deformation of the
collapsing core. The centrifugally-deformed core undergoes extreme
accelerations during the late collapse, bounce, and early postbounce
phase. This provides an extremely large accelerated quadrupole moment
(e.g.,~\cite{fryernew:11}), resulting in a GW burst signal that,
depending on the amount of angular momentum in the inner core, can be
detected by the upcoming advanced-generation of GW detectors out to
$10-100\,\mathrm{kpc}$
\cite{dimmelmeier:08,ott:12a,abdikamalov:10}. After core bounce, on a
timescale of tens of milliseconds, non-axisymmetric dynamics may
develop due to rotational shear instabilities (e.g.,
\cite{ott:07prl,scheidegger:10b,kuroda:13}), leading to longer-term
quasi-periodic GW emission. 

Much effort has gone into modeling the GW signal from rotating core
collapse and bounce over the past three decades
\cite{mueller:82,moenchmeyer:91,yamadasato:95,zwerger:97,kotake:03,dimmelmeier:02,shibata:04,ott:04,ott:07prl,ott:07cqg,dimmelmeier:07,dimmelmeier:08,takiwaki:11,ott:12a}
and the current state of the art is set by simulations in
conformally-flat or full general relativity (GR) that include
realistic EOS and approximate neutrino transport
\cite{ott:12a,dimmelmeier:08,abdikamalov:10,kuroda:13}. These studies
found that the GW signal from rapidly rotating core collapse and
bounce has rather simple morphology and can be described by a
prebounce rise in GW strain $h$, a large spike at bounce, and a
subsequent postbounce ring-down phase in which the protoneutron star
hydrodynamically dissipates its remaining pulsational energy from
bounce. Simulations that included magnetic fields showed that the
bounce and very early postbounce phase and the associated GW signal
are not affected by magnetohydrodynamic effects unless the precollapse
seed fields are unrealistically large ($B \gtrsim 10^{12}$)
\cite{kotake:04,obergaulinger:06a,obergaulinger:06b,shibata:06,takiwaki:11}.
Dimmelmeier~\emph{et al.}~\cite{dimmelmeier:08} and
Abdikamalov~\emph{et al.}~\cite{abdikamalov:10} showed that the peak
GW strain from collapse and bounce depends primarily and sensitively
on the mass and angular momentum of the inner core at
bounce. Dimmelmeier~\emph{et al.}~\cite{dimmelmeier:08}, who
considered two finite-temperature nuclear EOS, the EOS of
H.~Shen~\emph{et al.}~\cite{shen:98b,hshen:11} and the Lattimer-Swesty
EOS~\cite{lseos:91}, found only a weak dependence of the GW signal on
the nuclear EOS.  Ott~\emph{et al.}~\cite{ott:12a} recently showed
that in rapidly rotating cores that produce protoneutron stars with
spin periods $\lesssim 5\,\mathrm{ms}$, the GW signal depends on the
angular momentum of the precollapse core, but not on its detailed
structure and progenitor ZAMS mass. Furthermore, they demonstrated
that postbounce neutrino emission has little influence on the GW
signal from bounce and ring-down.

In this work, we extend previous studies and focus on the influence of
the angular momentum distribution in the progenitor core on the GW
signal of rotating core collapse, bounce, and ring-down. To this end,
we carry out 124 axisymmetric simulations with the GR core-collapse
code {\tt CoCoNuT}
\cite{dimmelmeier:02a,dimmelmeier:05,abdikamalov:10}. For collapse and
the very early postbounce phase, axisymmetry is an excellent
approximation (unless the inner part of the iron core contains large
nonaxisymmetric perturbations, which is unlikely;
cf. \cite{ott:07prl,ott:07cqg,kuroda:13}). We employ the
Lattimer-Swesty $K=220\,\mathrm{MeV}$ EOS \cite{lseos:91} and treat
electron capture during collapse with the deleptonization scheme of
\cite{liebendoerfer:05,dimmelmeier:08}. After bounce, we employ the
neutrino-leakage scheme used in \cite{ott:12a}. Motivated by the
findings of \cite{ott:12a}, we consider only a single progenitor model
(the $12$-$M_\odot$ [at ZAMS] solar-metallicity progenitor of
\cite{woosley:07}) and carry out a systematic set of simulations with
five different degrees of differential rotation and a fine-grained
grid of initial central angular velocities.  In order to understand
systematic uncertainties in our simulations, we explore their
dependence on the nuclear EOS and the electron fraction in the inner
core.

The results of our simulations show that the GW signal of rapidly
rotating cores has a strong and systematic dependence on the
precollapse degree of differential rotation in cores that collapse to
rapidly rotating protoneutron stars with ratio of rotational kinetic
energy to gravitational energy $\beta = T/|W| \gtrsim 0.08$.  The GW
signal of more slowly spinning cores has little dependence on
differential rotation and instead just depends on the core's ``total
rotation,'' parameterized by its $\beta = T/|W|$ at core bounce. We
supplement our simulation results with a matched-filtering analysis
and a Bayesian model selection analysis motivated by
\cite{logue:12}. Assuming Advanced LIGO (aLIGO; \cite{harry:10})
design sensitivity, we demonstrate that it is possible to measure
total rotation (i.e.\ $\beta$ or the angular momentum $J$) within
$\sim$20\% for unknown injected rotating core collapse signals from
galactic events, assuming optimal source-detector orientation and
Gaussian noise. We also show that the degree of differential rotation
can be estimated, but robustly only for rapidly rotating models and if
the EOS and inner-core electron fractoin are known.

This paper is organized as follows. In Section~\ref{sec:methods}, we
describe our computational code and in
Section~\ref{sec:initial_models}, we discuss our precollapse
configurations. Section~\ref{sec:results1} presents the results of our
core collapse simulations and analyzes the effects of differential
rotation on dynamics and GW signal. In Section~\ref{sec:gwanalysis},
we present the matched-filtering analysis and Bayesian model selection
results.  We summarize and conclude in Section~\ref{sec:summary}.


\section{Methods}
\label{sec:methods}

We perform our simulation in axisymmetric (2D) conformally-flat GR
with the {\tt CoCoNuT} code, which has been extensively described
in~\cite{dimmelmeier:02,dimmelmeier:05,dimmelmeier:08,abdikamalov:10}.
The conformal-flatness condition (CFC) has been shown to be an
excellent approximation to full GR in the context of rotating stellar
collapse to protoneutron stars~\cite{ott:07cqg}.  For the timescales
considered in this work, the small systematic errors due to CFC
approximation are completely dwarfed by the systematic uncertainties
associated with the nuclear EOS and the treatment of neutrinos.  {\tt
  CoCoNuT} employs Eulerian spherical coordinates and solves the
non-linear elliptic CFC equations using spectral
methods~\cite{dimmelmeier:05}.  GR hydrodynamics is implemented
following the Valencia formulation \cite{font:08} via a finite-volume
method with piecewise parabolic reconstruction~\cite{colella:84}, and
the approximate HLLE Riemann solver~\cite{HLLE:88}. The version of
{\tt CoCoNuT} used here is the same as in \cite{abdikamalov:10}, but
we have upgraded the EOS and neutrino microphysics routines as
described in the following.

We use the tabulated finite-temperature nuclear EOS by Lattimer \&
Swesty \cite{lseos:91} with $K=220\,\mathrm{MeV}$ generated
by~\cite{oconnor:10} and available for download from~{\tt
  stellarcollapse.org}. More information on this EOS and the details
of its implementation in a tabulated form can be found
in~\cite{oconnor:10,stellarweb}. In order to study the effect of the
nuclear EOS itself, we repeat a select set of our models with the
Shen~\emph{et~al.}~\cite{shen:98a,shen:98b} EOS, a table of which is
also available on {\tt stellarcollapse.org}.

We employ the neutrino microphysics routines provided by the
open-source code {\tt GR1D}~\cite{oconnor:10,oconnor:11,oconnor:11b},
also available for download from {\tt stellarcollapse.org}.  During
the collapse phase, we use the parameterized $Y_e(\rho)$
deleptonization scheme~\cite{liebendoerfer:05fakenu} with the same
parameters used in \cite{ott:12a} (see Appendix~\ref{sec:yepar} for
details). In the postbounce phase, we use the neutrino leakage/heating
scheme of \cite{oconnor:10} that approximates deleptonization,
neutrino cooling, and heating.  We implement the optical depth
calculation along radial rays aligned with {\tt CoCoNuT}'s radial
zones and use the default heating scaling factor $f_\mathrm{heat} = 1$
of this scheme. We take into account the contribution from neutrinos
to the hydrodynamic pressure and the spacetime stress-energy tensor in
the optically thick region via the ideal Fermi gas approximation above
a fiducial neutrino trapping density of $2\times
10^{12}\,\mathrm{g}\,\mathrm{cm}^{-3}$, following the prescription
of~\cite{liebendoerfer:05fakenu}. This leakage/heating scheme has also
been applied in the multi-dimensional simulations of
\cite{ott:12a,ott:13a,couch:13e}.

As in previous studies (e.g., \cite{dimmelmeier:08,abdikamalov:10}),
we perform our simulations in a spherical domain spanning 3000 km in
radius under the assumption of equatorial symmetry. In our production
simulations, we cover our domain with 250 logarithmically spaced
radial grid points with a central resolution of 250 m. The $90^\circ$
of our domain are covered with 40 equidistant angular grid points. We
have performed a resolution study to ensure that this resolution is
sufficient for the purpose of this study.

We extract GWs using the variant of the Newtonian quadrupole formula
given in \cite{dimmelmeier:05}, which is very accurate in the case of
rotating stellar collapse to protoneutron
stars~\cite{reisswig:11ccwave}. {More specifically, this method has
  negligible phase error, while the GW strain amplitudes are
  captured within a few per cent of their true values obtained in full
  GR simulations with Cauch-Characteristic Extraction
  \cite{reisswig:11ccwave}. This level of accuray is sufficient for
  studying how the waveforms depend on rotation and the degree of
  differential rotation, which is the main goal of this paper.


\section{Initial Models}
\label{sec:initial_models}

\begin{figure}[t]
\centering
\includegraphics[width=0.97\columnwidth]{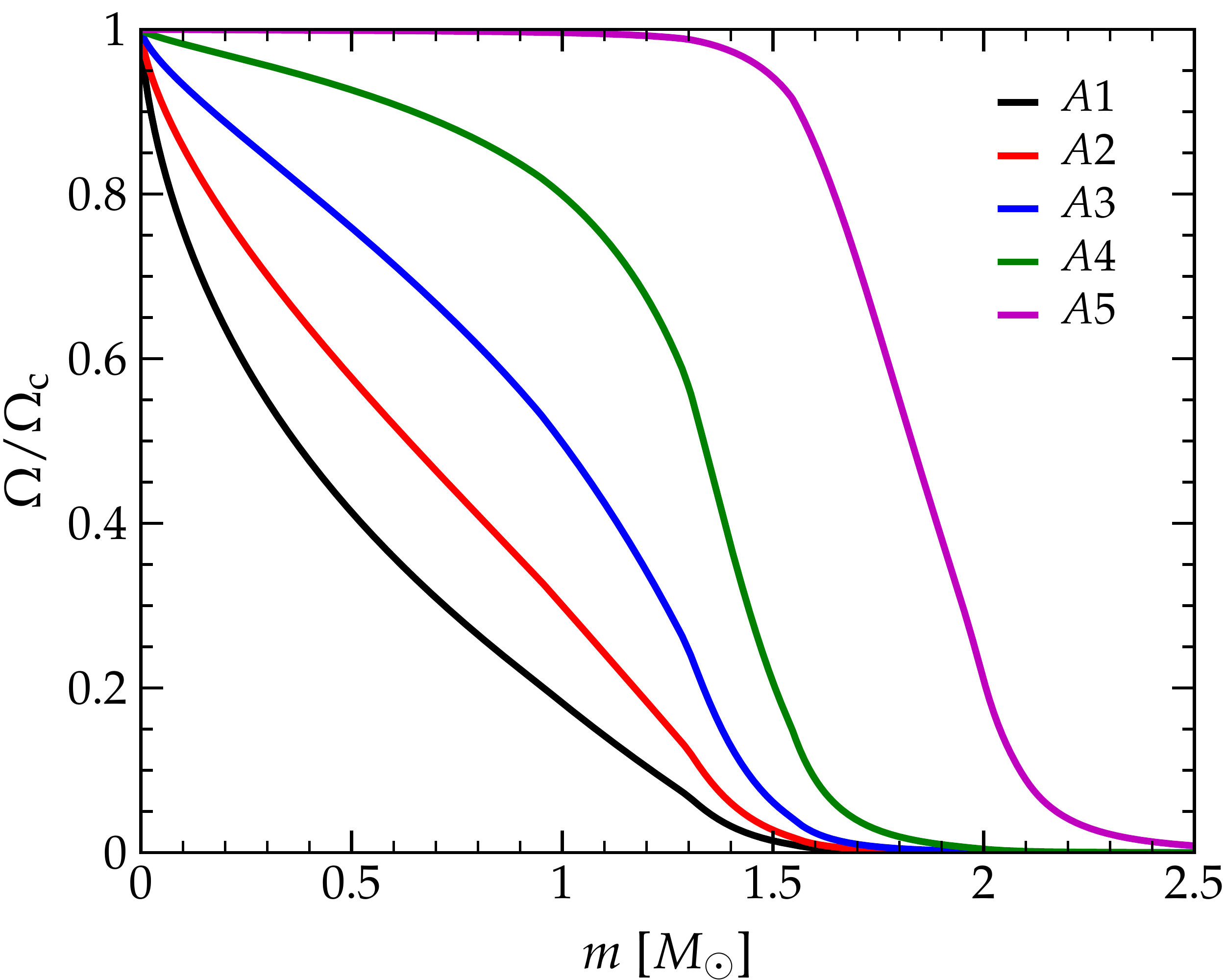}
\caption{The ratio of the angular velocity to the central angular
  velocity as a function of the enclosed-mass coordinate along the
  equatorial plane for the s12WH07 progenitor and for the five different
  values of the differential rotation parameter $A$ considered in this
study (cf.~Table~\ref{tab:models}).}
\label{fig:omega_vs_m}
\vspace{0.5ex}
\end{figure}

Existing presupernova stellar models with rotation are evolved using
spherically-symmetric codes assuming shellular rotation~(e.g.,
\cite{heger:00,heger:05,woosley:06}). In these models, the key
processes that determine the precollapse rotational configuration, such
as the magnetic braking~\cite[e.g.,][]{spruit:02} and mass
loss~\cite[e.g.,][]{langer:12}, are treated only approximately,
while the potentially important effects of binary
interactions~\cite[e.g.,][]{demink:13} are generally not
included at all.

Since our knowledge of the precollapse rotational configuration is far
from being certain, we employ the non-rotating $12$-$M_\odot$
solar-metallicity progenitor model of \cite{woosley:07} (model
s12WH07) and impose a simple parametrized rotation profile, which
facilitates control of the total angular momentum and its
distribution. We use the cylindrical rotation law
of~\cite{zwerger:97,ott:04}, 
\begin{equation}
\Omega(\varpi) = \Omega_\mathrm{c} \left[ 1 + \left(\frac{\varpi}{A}
  \right)^2 \right]^{-1}\,\,, 
\label{eq:rotlaw}
\end{equation}
where $\Omega_\mathrm{c}$ is the initial central angular velocity,
$\varpi$ is the cylindrical radius, and $A$ is the parameter that
controls the degree of differential rotation. This rotation law yields
constant specific angular momentum at $\varpi \gg
A$. Upon mapping into the code, the spherically
  symmetric initial model is set into rotation according to
  Eq.~(\ref{eq:rotlaw}). Collapse proceeds more slowly than the sound
  crossing time of the core and the latter is quickly driven into an
  oblate shape by centrifugal effects. The validity of this approach was studied
  by~\cite{zwerger:97}.

It is important to note that it is currently unclear how realistic the
rotation law given by Eq.~(\ref{eq:rotlaw}) is.  We use
  it nevertheless, since it represents the current standard way in
  which to setup rotating core collapse and because we
require a rotation law that (\emph{i}) roughly reproduces the angular
momentum distribution expected in stellar cores\footnote{The rotation
  law given by Eq.~(\ref{eq:rotlaw}) reproduces the radial angular
  momentum distribution in, e.g., rapidly rotating models 16TI and
  16OM of Woosley \& Heger \cite{woosley:06} with reasonable accuracy
  in the inner $\sim 2 M_\odot$ for $A\sim 850$ km.}, (\emph{ii}) does
not violate any known physical principles and constraints that are
relevant in this regime, and (\emph{iii}) allows us to easily
construct models with different amounts and distributions of angular
momentum. The rotation law given by Eq.~(\ref{eq:rotlaw}) fulfills
these requirements.

We restrict our analysis to a single progenitor model, since different
models with the same distribution of angular momentum as a function of
enclosed mass are likely to produce very similar dynamics and GW
signals at bounce and in the early postbounce ring-down phase. This
was demonstrated by \cite{ott:12a}.

We consider five sets of models with five different values of the
differential rotation parameter $A$: $A1 = 300\,\mathrm{km}$, $A2 =
417\,\mathrm{km}$, $A3 = 634\,\mathrm{km}$, $A4 =
1268\,\mathrm{km}$, $A5 = 10000\,\mathrm{km}$.
Figure~\ref{fig:omega_vs_m} depicts the ratio
$\Omega/\Omega_\mathrm{c}$ as a function of mass coordinate for these
values of $A$ in the s12WH07 progenitor model. The higher the value of
$A$, the weaker the differential rotation. The specific choices of
$Ai$ are motivated as follows: $A3$ is the same value used in
\cite{ott:12a} and gives an angular velocity at a mass coordinate of
$1\,M_\odot$ that is one half of the central value. $A4$ is twice as
large as $A3$, allowing us to probe somewhat more rigid initial
rotation, and $A5$ ensures near uniform rotation in the inner
$1.5\,M_\odot$ (corresponding to a radius of $\sim 3 \times
10^3\,\mathrm{km}$). $A1$ corresponds to extreme differential
rotation, and $A2$ is in the middle between $A1$ and $A3$.

For each choice of $A$, we simulate sequences of models with initial
central angular velocities starting at $\Omega_\mathrm{c,min} =
1\,\mathrm{rad}\,\mathrm{s}^{-1}$ (for this value, rotation is
dynamically insignificant in all models). We increase
$\Omega_\mathrm{c}$ in steps of
$0.5\,\mathrm{rad}\,\mathrm{s}^{-1}$. In models with weak or moderate
differential rotation (sequences $A3-A5$) the maximum initial central
angular velocity $\Omega_\mathrm{c,max}$ is set by the value at which
such models still collapse. For more differentially rotating models,
we choose $\Omega_\mathrm{c,max}$ in such a way that we obtain the
global maximum of $\beta_\mathrm{ic,b} = (T/|W|)_\mathrm{ic,b}$, the
ratio of rotational kinetic energy to gravitational energy of the
inner core at bounce. We compute $T/|W|$ via the definition given by
\cite{friedman:86} and focus on the inner core, because the bounce
dynamics and the associated GW signal are determined by its spin and
mass \cite{dimmelmeier:08,abdikamalov:10}. Note that models with
$\Omega_\mathrm{c} > \Omega_\mathrm{c,max}$ yield decreasing
$\beta_\mathrm{ic,b}$ (see, e.g., the discussions in
\cite{ott:04,ott:06spin,dimmelmeier:08}).  Since such models collapse
only in the case of very strong differential rotation, they are not
useful for our goal of comparing different degrees of differential
rotation.

\begin{table*}[t]
\begin{center}
\begin{tabular}{lccccccc}
\hline
\hline
Model    & $A$  &$\Omega_\mathrm{c,min}$ & $\Omega_\mathrm{c,max}$ & $\beta_\mathrm{ic,b,min}$ & $\beta_\mathrm{ic,b,max}$ & Number\\
sequence & [km] &[$\mathrm{rad\,s^{-1}}$] & [$\mathrm{rad\,s^{-1}}$] & [$10^{-2}$]& & of models \\
\hline
$A1$ & 300  & 1 & 15.5\z& 1.62 & 0.21 & 30\\
$A2$ & 417  & 1 & 11.5\z& 3.13 & 0.19 & 22\\
$A3$ & 634  & 1 & 9.5   & 3.58 & 0.18 & 18\\
$A4$ & 1268 & 1 & 6.5   & 4.66 & 0.13 & 12\\
$A5$ & 10000& 1 & 5.5   & 5.15 & 0.11 & 10\\
\hline\hline
\end{tabular}
\caption{Summary of key parameters of our model
  sequences. $\Omega_\mathrm{c,max}$ is the central angular velocity
  corresponding to the fastest spinning model in each
  $A$-sequence. $\beta_\mathrm{ic,b,min}$ and
  $\beta_\mathrm{ic,b,max}$ are the values of $\beta = T/|W|$ of the
  inner core at bounce for the slowest and fastest rotators of each
  sequence, respectively. Note that $\Omega_\mathrm{c,max}$ and
  $\beta_\mathrm{ic,b,max}$ in the only mildly differentially rotating
  sequence $A4$ and $A5$ are limited by the fact that more rapidly
  spinning models fail to collapse. In more differentially rotating
  models, $\Omega_\mathrm{c,max}$ is the value for which we obtain
  $\beta_\mathrm{ic,b,max}$. Due to centrifugal effects, models with
  higher initial $\Omega_\mathrm{c}$ yield smaller
  $\beta_\mathrm{ic,b}$ (see, Fig.~\ref{fig:omicb_jicb_vs_betaicb},
  and, e.g., \cite{ott:04,ott:06spin,dimmelmeier:08}).}
\label{tab:models}
\end{center}
\end{table*}


\section{Results: Dynamics and Waveforms}
\label{sec:results1}

\begin{figure}[t]
\centering
\includegraphics[width=0.97\columnwidth]{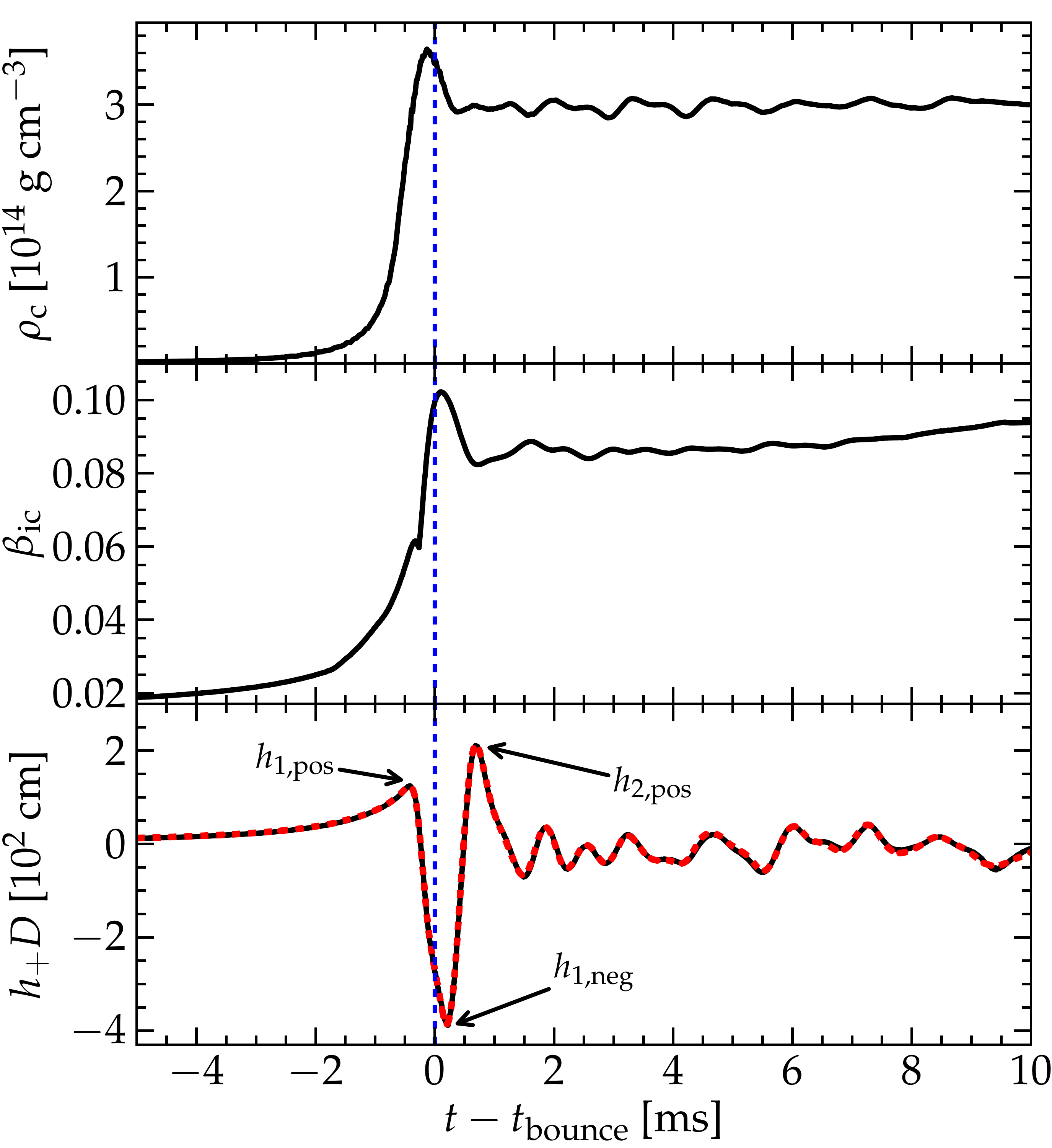}
\caption{Time evolution of the central density (top panel),
  $\beta_\mathrm{ic}$ (center panel), and GW strain (bottom panel;
  rescaled by source distance $D$) in model $A3O6$. The arrows indicate
  the first three pronounced generic features of the GW signal,
  labeled $h_{1,\mathrm{pos}}$, $h_{1,\mathrm{neg}}$, and
  $h_{2,\mathrm{pos}}$. The thin vertical dashed line indicates the
  time of core bounce defined as the time at which the equatorial edge
  of the inner core reaches an entropy of
  $3\,k_\mathrm{B}\,\mathrm{baryon}^{-1}$. The dashed red line shows
  the GW strain for the same model simulated with $50\,\%$ higher
  resolution in both the angular and radial direction. There is excellent
  agreement, which suggests that our fiducial resolution yields
  converged results.}
\label{fig:rho_gwstrain_beta_vs_t}
\vspace{0.5ex}
\end{figure}

\begin{figure}[t]
\centering
\includegraphics[width=0.97\columnwidth]{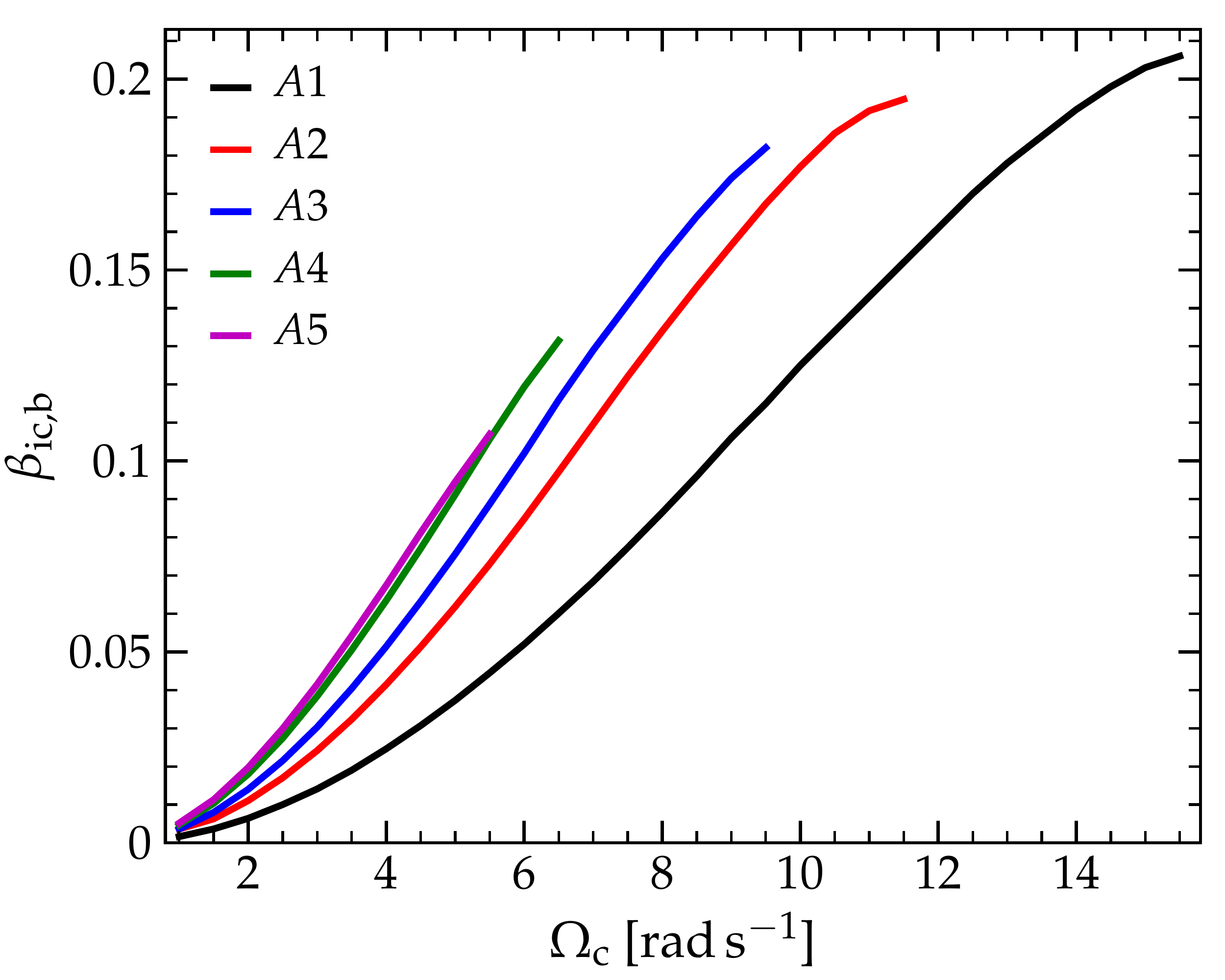}
\caption{Ratio of rotational kinetic energy to gravitational
  energy of the inner core at bounce $\beta_\mathrm{ic,b}$ as a
  function of initial central angular velocity
  $\Omega_\mathrm{c}$. All model sequences, from near uniform rotation
  ($A5$) to strong differential rotation ($A1$), are
  shown. Sequences with uniform or moderate differential rotation
  terminate at $\Omega_c$ beyond which they would be fully
  centrifugally supported already at the onset of
  collapse. Cf.~Table~\ref{tab:models}. Note that the mapping
  $\Omega_c \rightarrow \beta_\mathrm{ic,b}$ depends on progenitor
  structure \cite{ott:12a}.}
\label{fig:betaicb_vs_omegac}
\vspace{0.5ex}
\end{figure}

The top panel of Fig.~\ref{fig:rho_gwstrain_beta_vs_t} depicts the
time evolution of the central density $\rho_\mathrm{c}$ during the
last phase of collapse, bounce, and the early postbounce phase of
model $A3O6$, which is representative for many of the simulated
models. For future reference, we define the \emph{time of core bounce}
as the moment at which the specific entropy at the edge
of the inner core in the equatorial plane reaches
$3\,k_\mathrm{B}\,\mathrm{baryon}^{-1}$. Just before 
bounce, $\rho_\mathrm{c}$ increases rapidly due to the accelerated
contraction of the inner core. Once nuclear density is reached, the
stiffening of the nuclear EOS abruptly decelerates collapse. The inner
core overshoots its equilibrium configuration due to its immense
inertia, and consequently $\rho_\mathrm{c}$ reaches $\sim\! 3.7 \times
10^{14}\,\mathrm{g\,cm}^{-3}$ at maximum contraction. The core bounces
back and settles at a postbounce (pb) quasi-equilibrium central
density $\rho_\mathrm{c,pb}$ of $\sim\! 3 \times
10^{14}\,\mathrm{g\,cm}^{-3}$, after a series of ring-down
oscillations that last for $\sim\! 10-15$ ms. These oscillations are
clearly visible in the evolution of the central density in the
postbounce phase as a quasi-periodic $\sim\! 7 \, \%$ variation of
$\rho_\mathrm{c}$.

The middle panel of Fig.~\ref{fig:rho_gwstrain_beta_vs_t} shows $\beta
= T/|W|$ of the inner core for model $A3O6$. By construction, $\beta$
directly reflects the importance of rotational support in
gravitationally bound objects~\cite{tohline:84}.  $\beta_\mathrm{ic}$
grows in the final phase of collapse due to the spin-up of the inner
core, as a consequence of angular momentum conservation. At bounce,
$\beta_\mathrm{ic}$ peaks at $\sim 0.1$ in this model, before
decreasing to $\sim 0.08$ while the inner core settles into its
postbounce quasi-equilibrium. The ring-down protoneutron star
oscillations in the postbounce phase are also visible as small
variations in $\beta_\mathrm{ic}$.

In Fig.~\ref{fig:betaicb_vs_omegac}, we show $\beta_\mathrm{ic,b}$ as
a function of the initial central angular velocity $\Omega_\mathrm{c}$
for all models. At fixed $\Omega_\mathrm{c}$, strongly differentially
rotating models reach smaller $\beta_\mathrm{ic,b}$ than more
uniformly rotating ones. This is due to the comparatively modest total
angular momentum in the former's inner cores.  With increasing
$\Omega_\mathrm{c}$, uniformly and mildly differentially rotating
models (sequences $A3-A5$) eventually become fully centrifugally
supported at the start of the simulation and do not collapse. Due to
this, the graphs in Fig.~\ref{fig:betaicb_vs_omegac} for such models
terminate at small to moderate $\Omega_c$, and the corresponding
maximum $\beta_\mathrm{ic,b}$ reached for sequences $A5$, $A4$, and
$A3$ are $0.11$, $0.13$, and $0.18$ respectively.

More strongly differentially rotating models, however,
collapse even at high $\Omega_c$. Sequences $A2$ and $A1$ reach
$\beta_\mathrm{ic,b}$ of 0.19 and 0.21, respectively. The graphs corresponding
to these models in Fig.~\ref{fig:betaicb_vs_omegac} show that these
values are very close to the obtainable global maximum.
A further increase in $\Omega_c$ would lead to a decrease in
$\beta_\mathrm{ic,b}$, because bounce occurs centrifugally at lower
core densities, corresponding to a smaller degree of spin-up (see the extensive
discussions in \cite{ott:04,ott:06spin,abdikamalov:10}).  


\subsection{Influence of Differential Rotation on Collapse, Bounce and Early Postbounce Dynamics}
\label{sec:dynamics}

\begin{figure}[t]
\centering
\includegraphics[width=0.97\columnwidth]{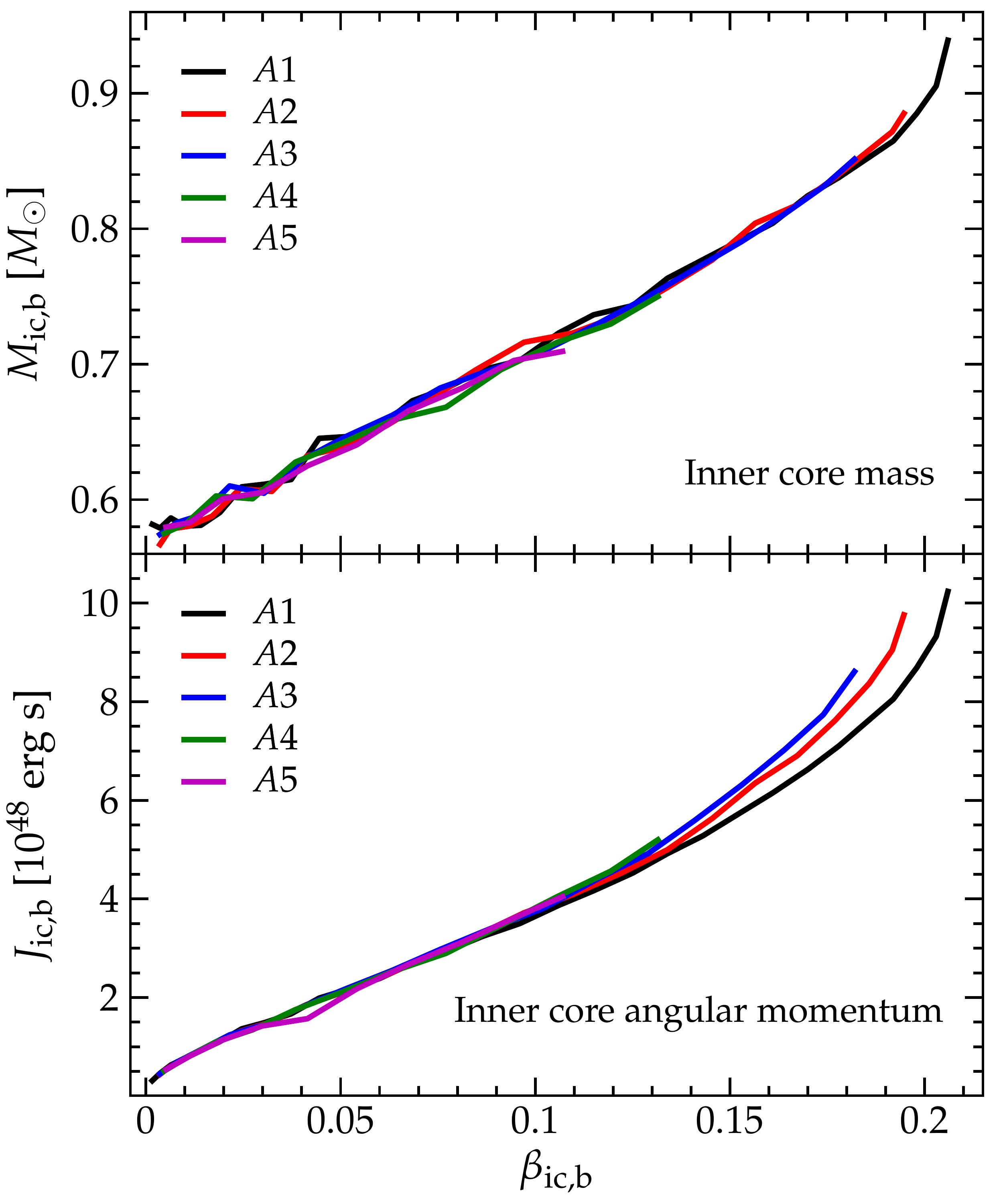}
\caption{Mass of the inner core at bounce ($M_\mathrm{ic,b}$, top
  panel) and angular momentum ($J_\mathrm{ic,b}$, bottom panel) as 
  functions of $\beta_\mathrm{ic,b}$ for all model sequences, varying
  from near uniform rotation ($A5$) to strong differential rotation
  ($A1$). $J_\mathrm{ic,b}$ increases linearly with
  $\beta_\mathrm{ic,b}$ and, for $\beta_\mathrm{ic,b} \lesssim 0.12$,
  is nearly independent of the degree of differential
  rotation. $M_\mathrm{ic,b}$ also increases with
  $\beta_\mathrm{ic,b}$ (and $J_\mathrm{ic,b}$) and is essentially
  independent of differential rotation for $\beta_\mathrm{ic,b}
  \lesssim 0.18$.} 
\label{fig:omicb_jicb_vs_betaicb}
\vspace{0.5ex}
\end{figure}

The central objective of this work is to infer the effects of the
angular momentum distribution in the progenitor core on the dynamics
of core collapse, bounce, the early ring-down oscillations, and the
resulting GW signal. As demonstrated already in previous work (e.g.,
\cite{dimmelmeier:08,dimmelmeier:07}), the effect on the inner core is
most important, since its mass and angular momentum (and, perhaps, its
distribution) determine the GW signal. We will discuss the
details of the latter in the next Section~\ref{sec:gw_peaks}.

As a start, it is useful to define a quantity that describes the
``total rotation'' of the inner core. One possibility is to use the
already introduced quantity $\beta_\mathrm{ic}$. It is most useful to
consider the value of $\beta_\mathrm{ic}$ at bounce, since this is
also the time at which the highest GW amplitudes occur. An obvious
alternative choice is the total angular momentum of the inner core
$J_\mathrm{ic}$, which, again, is best considered at the time of core
bounce. Another alternative, though less direct measure is the mass of the
unshocked inner core at bounce $M_\mathrm{ic,b}$. In the nonrotating
case, $M_\mathrm{ic,b}$ is determined by the trapped lepton fraction in
the inner core (e.g., \cite{bethe:90}). Rotation increases
$M_\mathrm{ic,b}$ by slowing down collapse and thus allowing a greater
amount of material to be in sonic contact and part of the inner core
\cite{moenchmeyer:91,dimmelmeier:08,abdikamalov:10}.

Figure~\ref{fig:omicb_jicb_vs_betaicb} shows that $\beta_\mathrm{ic,b}$,
$J_\mathrm{ic,b}$, and $M_\mathrm{ic,b}$ obey a simple linear
relationship and are independent of the degree of differential
rotation through most of the considered model parameter space.  Thus
they can be used interchangeably to describe ``total rotation''.
The simple relationship becomes non-linear and dependent on the
differential rotation parameter $A$   only for
very rapid rotation ($\beta_\mathrm{ic,b} \gtrsim 0.13$,
$J_\mathrm{ic,b} \gtrsim 6\times 10^{48}\,\mathrm{erg\,s}$,
$M_\mathrm{ic,b} \gtrsim 0.8\,M_\odot$). 

The mapping $\beta_\mathrm{ic,b} \rightarrow J_\mathrm{ic,b}$, shown
in the lower panel of Fig.~\ref{fig:omicb_jicb_vs_betaicb}, exhibits
interesting dependence on $A$ in rapidly rotating models with
$\beta_\mathrm{ic,b}\gtrsim 0.13-0.15$. More differentially rotating
models have systematically less $J_\mathrm{ic,b}$ at fixed
$\beta_\mathrm{ic,b}$ than less differentially rotating ones. This is
straightforward to understand, since, at fixed $J_\mathrm{ic,b}$ and
$M_\mathrm{ic,b}$, a more differentially rotating inner core will
always have more rotational energy. Hence, at fixed
$\beta_\mathrm{ic,b}$ and $M_\mathrm{ic,b}$, $J_\mathrm{ic,b}$ for a
model with smaller $A$ will be smaller.

\begin{figure}[t]
\centering
\includegraphics[width=0.97\columnwidth]{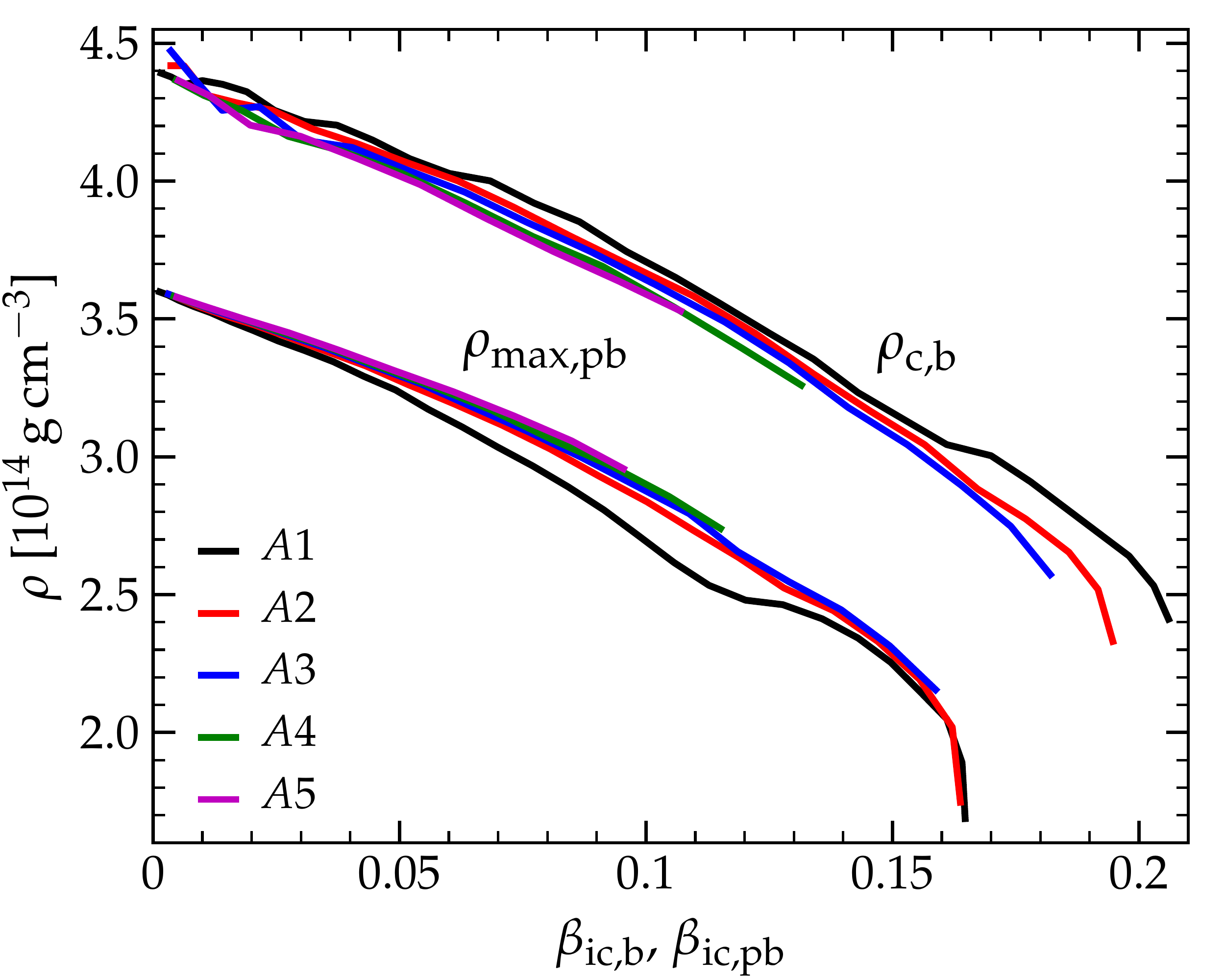}
\caption{The central rest-mass density at bounce $\rho_\mathrm{c,b}$
  as a function of $\beta_\mathrm{ic,b}$ (upper graphs) and
  time-average maximum density in the postbounce phase
  $\rho_\mathrm{c,pb}$ as a function of the time-averaged
  $\beta_\mathrm{ic,pb}$ (lower graphs).  We show curves for the five
  values of the differential rotation parameter $A$. Centrifugal
  support leads to a decrease in the density both at bounce and in
  the postbounce core. A strong dependence on differential rotation is
  apparent only in very rapidly rotating models. Note that
  differentially rotating models develop slightly off-center density
  maxima and quasi-toroidal structure
  (cf.~Fig.~\ref{fig:ent_colormap}), but the maximum density
  exceeds the central density only by a few percent in such models.}
\label{fig:rho_vs_betaicb}
\vspace{0.5ex}
\end{figure}

The central rest-mass density is important for the structure and
dynamics of the inner core, which turns into the unshocked
protoneutron star core after bounce. In the nonrotating,
low-temperature limit, the central density, for a given nuclear EOS,
determines stellar structure and pulsational mode spectrum completely
\cite{shapteu:83}.  In Fig.~\ref{fig:rho_vs_betaicb}, we plot the
central density at bounce ($\rho_\mathrm{c,b}$ as a function of
$\beta_\mathrm{ic,b}$; upper graphs) and the time-averaged density
over the first few milliseconds after bounce ($\rho_\mathrm{max,pb}$ as
a function of $\beta_\mathrm{ic,pb}$; lower graphs; we average over
$6\,\mathrm{ms}$, from $2$ to $8\,\mathrm{ms}$ after bounce). Both
quantities decrease with increasing total rotation, since centrifugal
support keeps the core in a less compact (i.e.\ lower-density)
configuration. The central densities of very slowly rotating models
($\beta_\mathrm{ic,b} \lesssim 0.02-0.03$) exhibit little variation
with differential rotation parameter $A$. In more rapidly rotating
models, those with smaller $A$ (more differential rotation) have
systematically slightly larger $\rho_\mathrm{c,b}$. Since most of
their spin is concentrated at small radii (and mass coordinates), they
experience less centrifugal support throughout the collapsing inner
core than models with larger $A$ at the same
$\beta_\mathrm{ic,b}$. However, after bounce the extremely rapid
rotation in the central regions of strongly differentially spinning
models leads to slightly more oblate innermost cores and somewhat
lower time-average postbounce densities, as shown by
Fig.~\ref{fig:rho_vs_betaicb}.

\begin{figure}[t]
\centering
\vspace*{-0.6cm}
\begin{minipage}{\linewidth}
\includegraphics[width=0.97\linewidth]{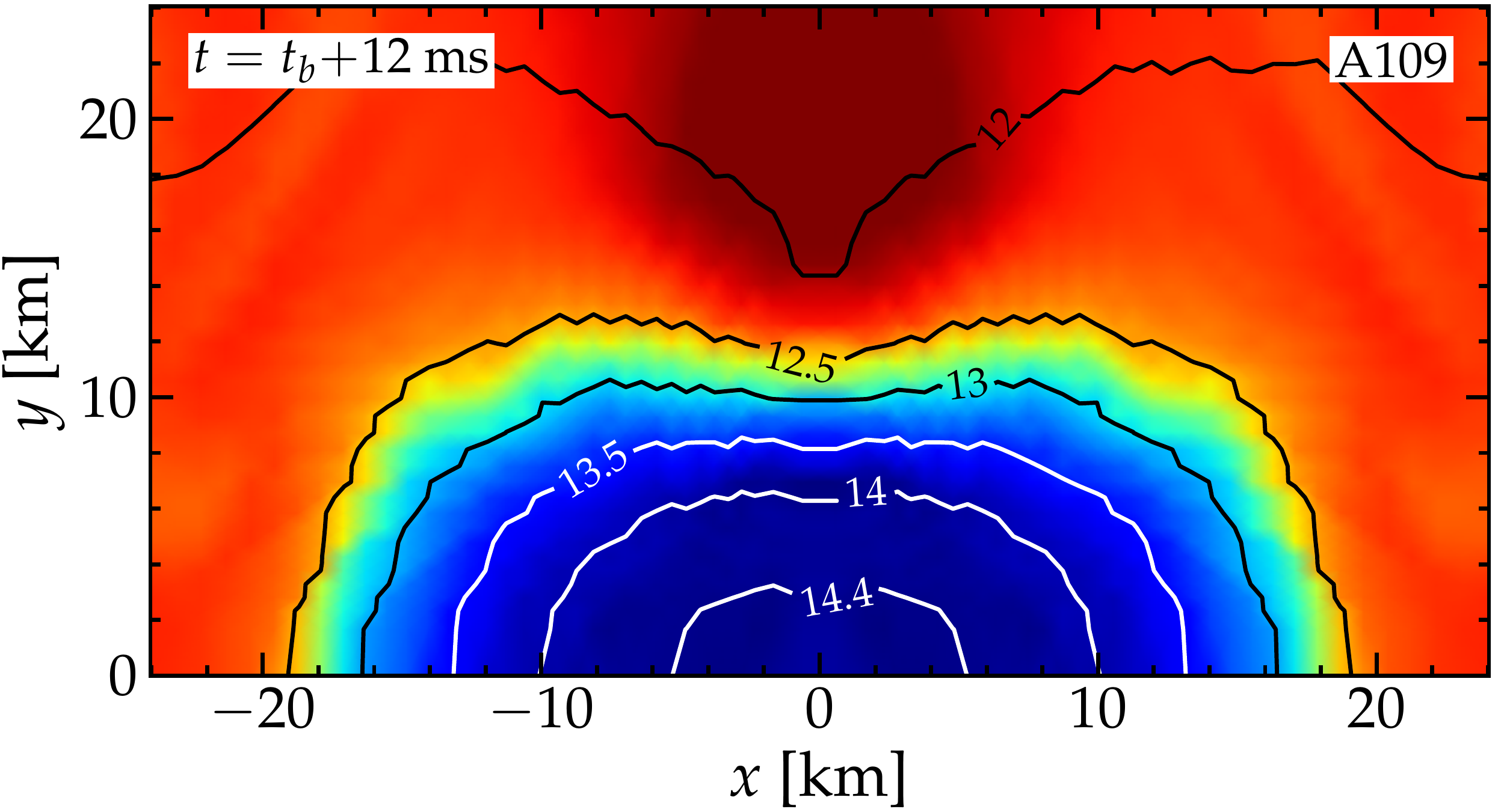}
\end{minipage}
\begin{minipage}{\linewidth}
\vspace*{-0.0cm}
\includegraphics[width=0.97\linewidth]{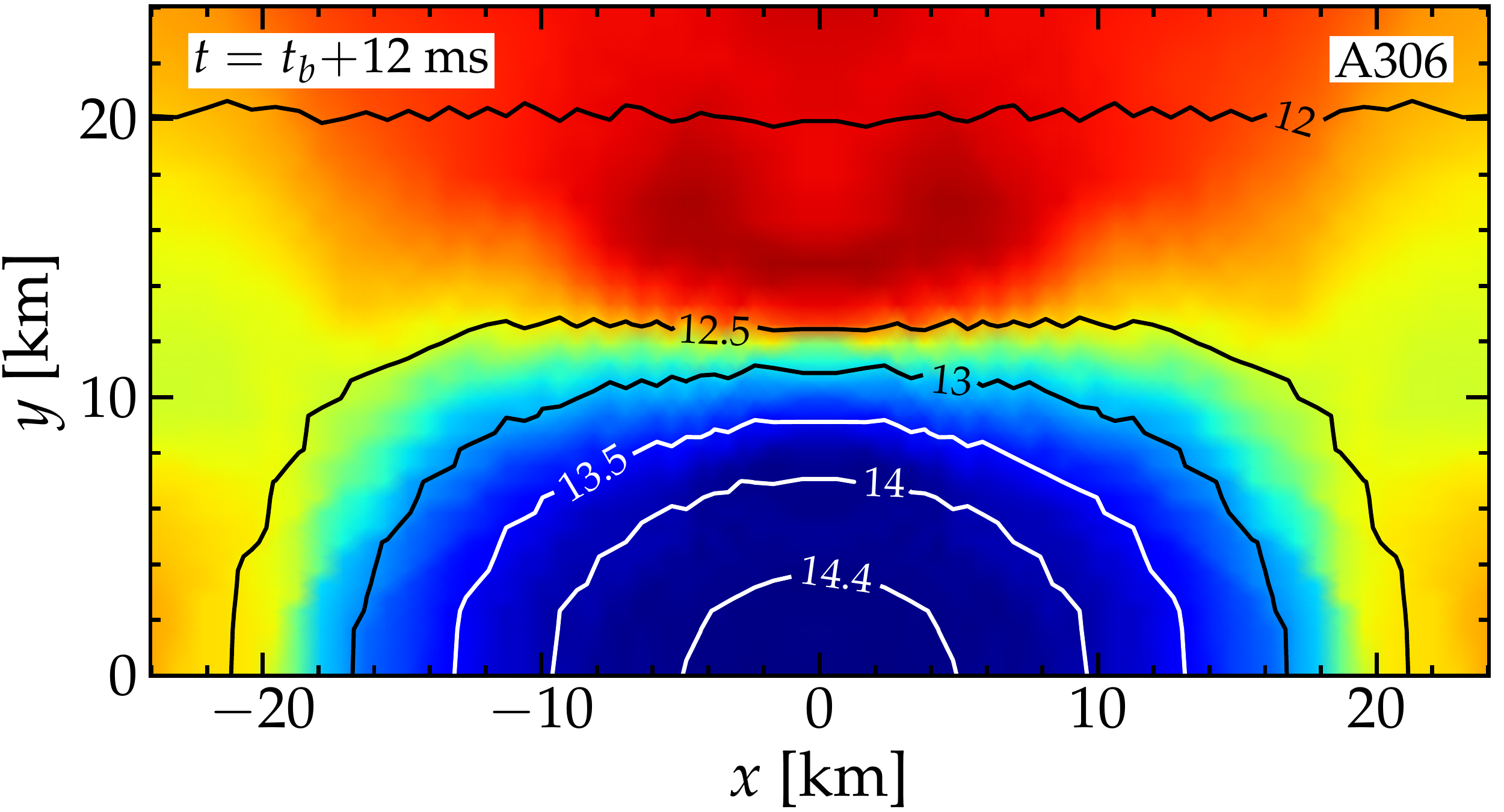}
\end{minipage}
\begin{minipage}{\linewidth}
\vspace*{-0.0cm}
\includegraphics[width=0.97\linewidth]{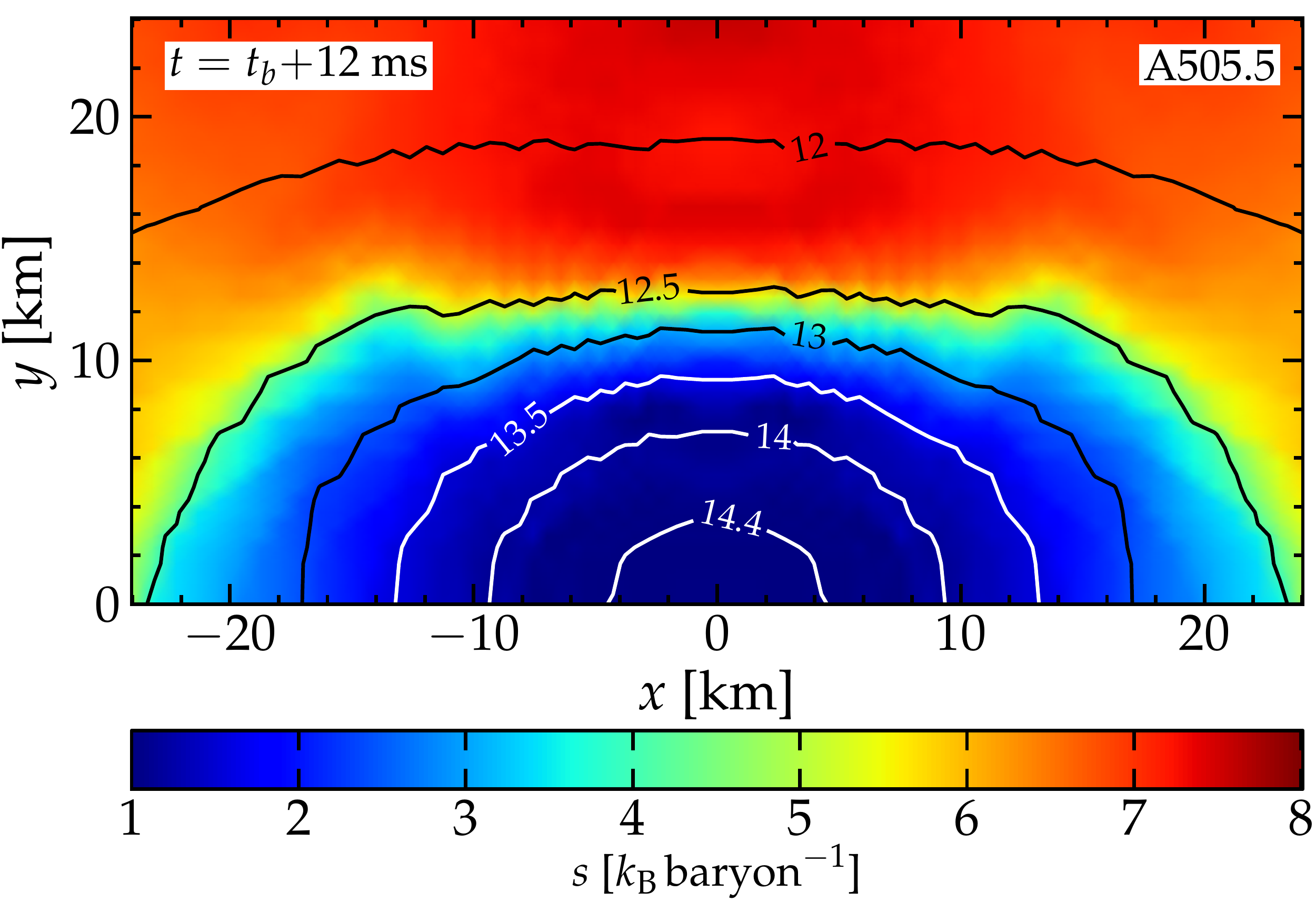}
\end{minipage}
\caption{Entropy colormaps of the meridional plane for models $A1O9$,
  $A3O6$, $A5O5.5$ with $\beta_\mathrm{ic,b}\,\sim\,0.1$ at
  $12\,\mathrm{ms}$ after bounce. Black and white lines are mark
  density isocontours at $10^{12}$, $10^{12.5}$, $10^{13}$,
  $10^{13.5}$, $10^{14}$, and $10^{14.4}\,\mathrm{g\,cm^{-3}}$. More
  differentially rotating models have more compact unshocked (low
  entropy) cores and more centrifugally deformed innermost density
  isocontours.}
\label{fig:ent_colormap}
\end{figure}

Figure~\ref{fig:ent_colormap} depicts 2D entropy colormaps with
superposed iso-density contours at $12\,\mathrm{ms}$ after bounce for
three representative models with $\beta_\mathrm{ic,b} \sim 0.1$ and
differential rotation parameters $A1$ (model $A1O9$, strong
differential rotation), $A3$ (model $A3O6$, moderate differential
rotation), and $A5$ (model $A5O5.5$, nearly uniform rotation). Shown
are the upper hemisphere and the rotation axis is aligned with the
positive $z$-axis. The unshocked protoneutron star core (specific
entropy $s \lesssim 3\,k_\mathrm{B}\,\mathrm{baryon}^{-1}$) is more
extended in less differentially rotating models, since these have more
angular momentum at larger mass (and radial) coordinate. 

Figure~\ref{fig:ent_colormap} also shows that overall shape of
the protoneutron star cores varies with differential rotation. While
the $A5$ model is clearly spheroidal, the density contours (traced by
the entropy distribution) of the strongly differentially rotating
$A1$ model show a double-lobed structure characteristic of
quasitoroidal equilibrium configurations that have their maximum
density not at a single point at the origin, but in a ring at some
finite radius in the equatorial plane. This is expected to occur in
strongly differentially rotating cores and has been reported before
in, e.g.,
\cite{zwerger:97,dimmelmeier:02,dimmelmeier:08,abdikamalov:10}. We
indeed find that the tendency to develop off-center density maxima
increases with decreasing $A$, but even the most rapidly
differentially rotating model in our entire set has a density contrast
of only $\rho_\mathrm{max,pb} / \rho_\mathrm{c,pb} - 1 \lesssim 3\%$,
and its density maximum is located only $\sim$$1.2\,\mathrm{km}$ off
the origin.

\begin{figure}[t]
\centering
\includegraphics[width=0.97\columnwidth]{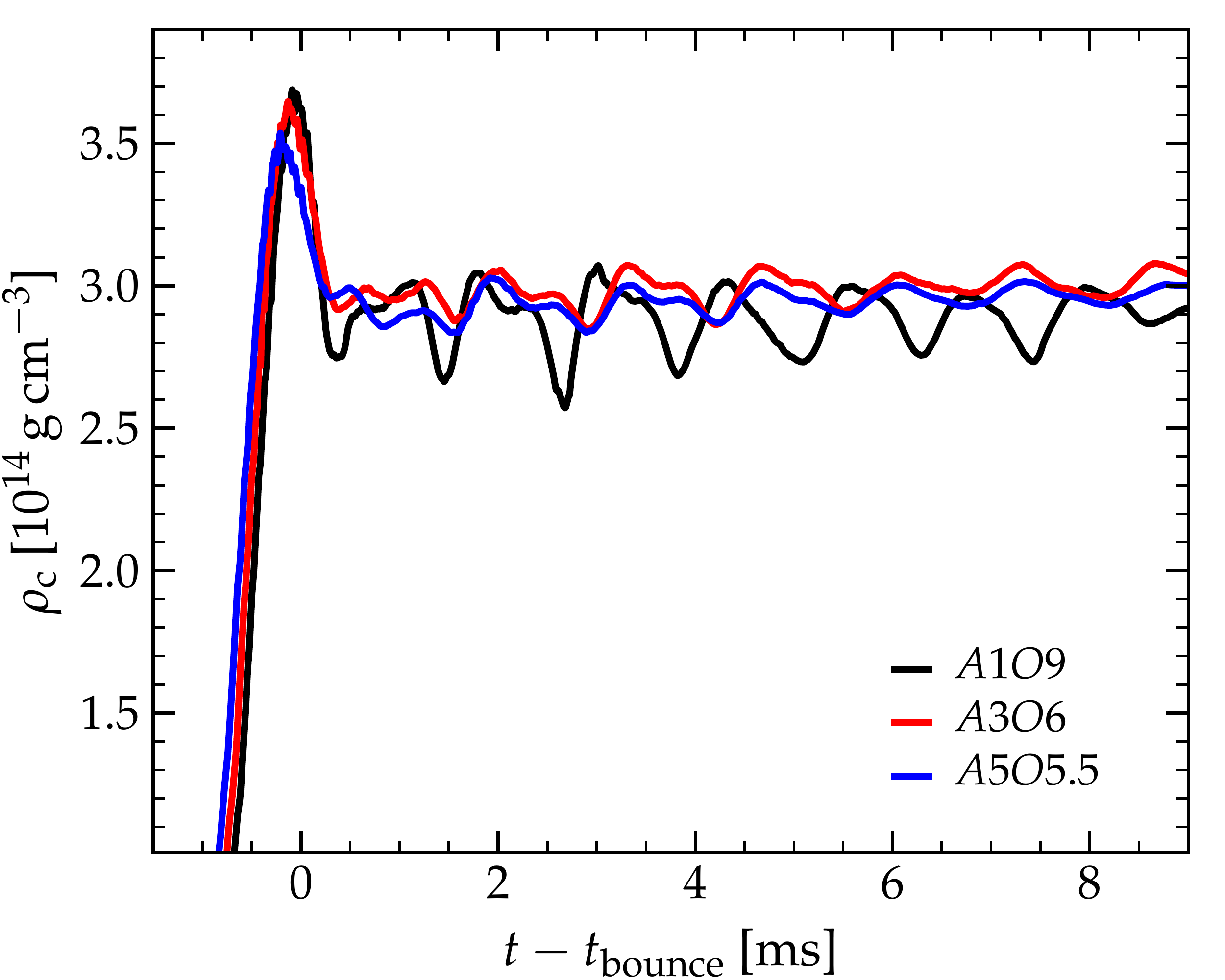}
\caption{Time evolution of the central density for models $A1O9$,
  $A3O6$, and $A5O5.5$, all of which have $\beta_\mathrm{ic,b} \sim
  0.1$. More differentially rotating models more strongly overshoot their
  postbounce quasi-equilibrium central densities at bounce and exhibit
  stronger postbounce ring-down oscillations.
}
\label{fig:rho_vs_t}
\vspace{0.5ex}
\end{figure}

In Fig.~\ref{fig:rho_vs_t}, we plot the evolution of the central
density of the same three models ($A1O9$, $A3O6$, and $A505.5$) with
$\beta_\mathrm{ic,b} \sim 0.1$ shown in
Fig.~\ref{fig:ent_colormap}. At bounce, the most differentially
rotating model overshoots its postbounce quasi-equlibrium the most,
settles at the lowest $\rho_\mathrm{c,pb}$, and exhibits the strongest
postbounce ringdown oscillations. These oscillations are non-linear
and a superposition of multiple modes, but in previous work, at least
one of the modes has been identified as the fundamental quadrupole
mode of the protoneutron star core \cite{ott:12a}. The most
differentially rotating model has most of its spin concentrated in the
innermost regions. Hence, these regions are most oblate ($\ell = 2$)
in this model, yielding the strongest excitation of the quadrupole
core pulsation mode.

In summary and to connect to the next Section on GW emission: although
the important quantities $\rho_\mathrm{c,b}$ and $\rho_\mathrm{max\,
  or\, c,pb}$ depend primarily on $\beta_\mathrm{ic,b}$, we also
observe a dependence on the differential rotation parameter $A$, in
particular in rapidly rotating cases. This and the obvious differences
in the 2D structure of the postbounce cores shown in
Fig.~\ref{fig:ent_colormap} suggest that the detailed
multi-dimensional dynamics of the GW-emitting inner core are governed
not only by its total rotation, but also by the distribution of
angular momentum. We shall next investigate the effect of differential
rotation on the GW signal.


\subsection{Influence of Differential Rotation on the Gravitational Wave Signal}
\label{sec:gw_peaks}

\begin{figure}[t]
\centering
\includegraphics[width=0.97\columnwidth]{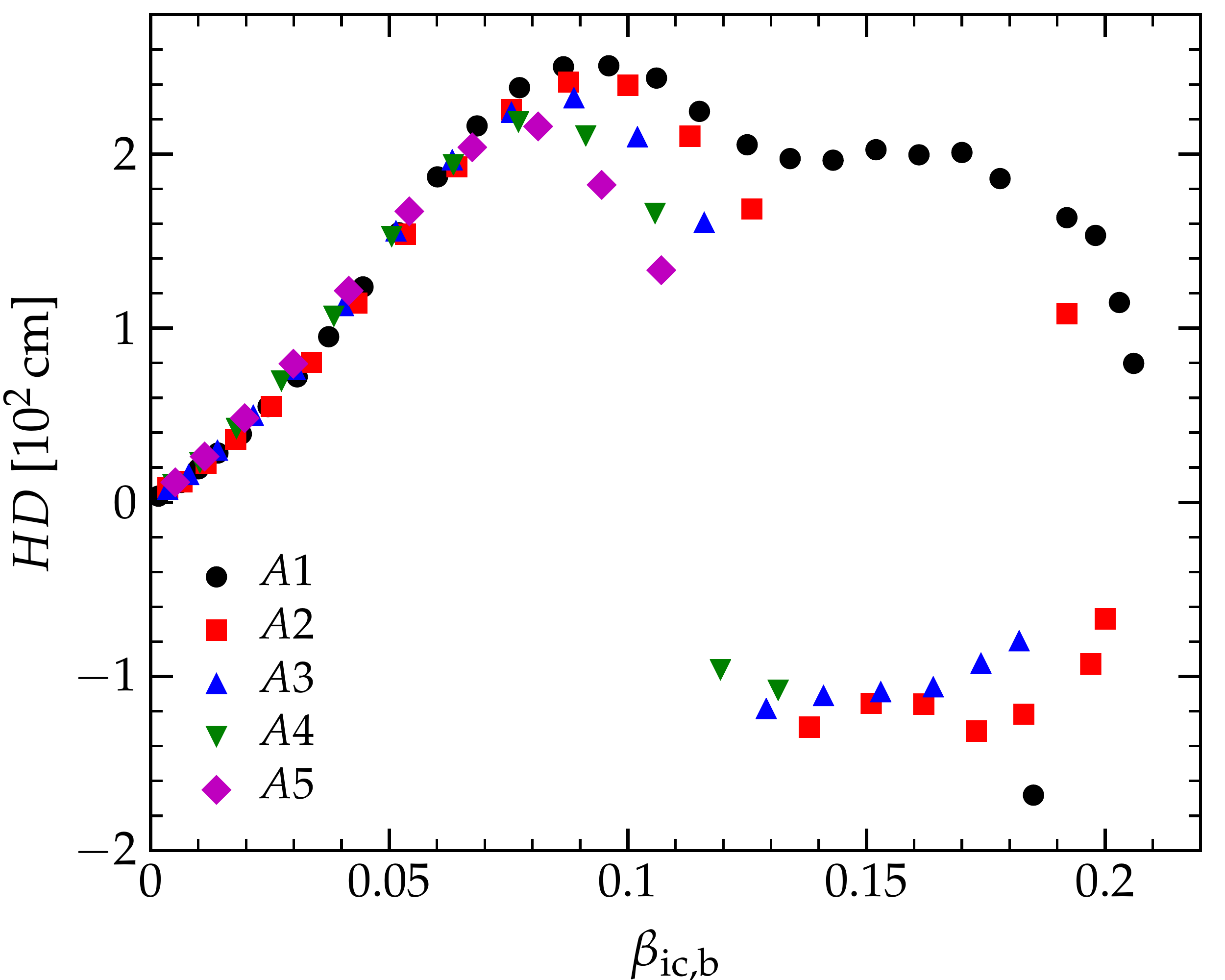}
\caption{Values of the second largest peak (in absolute value) ${H}$
  of the GW signal as a function of $\beta_\mathrm{ic,b}$ for  all
  models. More differentially rotating models yield larger positive
  ${H}$ and switch to negative ${H}$ at higher $\beta_{ic,b}$. While
  differential rotation is a necessary criterion for ${H} < 0$, it is
  not a sufficient one.} 
\label{fig:hpeak2_vs_betaicb}
\vspace{0.5ex}
\end{figure}

For an analysis of the influence of the differential rotation
parameter $A$ on the GW signal, it is useful to first recap the
latter's general morphology and at which point in the highly dynamical
evolution of the inner core it reaches its peak values. In the
following, without loss of generality, we will assume that the core's
spin is aligned with the positive $z$-axis. The bottom panel of
Fig.~\ref{fig:rho_gwstrain_beta_vs_t} shows the GW strain $h_+$ (there
is only one polarization due to axisymmetry) as a function of time in
the late collapse, bounce, and early postbounce phases of our
reference model $A3O6$. During the collapse phase, $h$ increases slowly
and reaches a positive peak, $h_\mathrm{1,pos}$, during the rapid
contraction phase immediately before bounce. During bounce, $h$
decreases rapidly, reaching its most pronounced negative peak
$h_\mathrm{1,neg}$ when the inner core is expanding at bounce (cf.~the
evolution of the maximum density shown in the top panel of
Fig.~\ref{fig:rho_gwstrain_beta_vs_t}). Following $h_\mathrm{1,neg}$,
$h$ reaches positive values and generically has a new positive local
maximum, $h_\mathrm{2,pos}$. In slowly rotating models
($\beta_\mathrm{ic,b} \lesssim 0.05$), $h_\mathrm{2,pos}$
coincides with the first recontraction of the core after bounce.  In
the rapidly rotating case, an identification of $h_\mathrm{2,pos}$
with global core dynamics is less obvious, since bounce leads to the
excitation of several oscillation modes in the core (dominated by the
fundamental quadrupole mode, see \cite{ott:12a}), which all contribute
to the GW signal at this point. After $h_\mathrm{2,pos}$, the core
undergoes ring-down oscillations that are damped hydrodynamically.
They produce more peaks in $h$ whose amplitudes decay on a timescale
of $10-15\,\mathrm{ms}$. Hereafter, we refer to the peaks that
occur after $h_{1,\mathrm{neg}}$ as \emph{ring-down peaks}.

Hayama~\emph{et~al.}~\cite{hayama:09} analyzed 2D Newtonian
simulations of 12 models simulated by
Kotake~\emph{et~al.}~\cite{kotake:03} with varying rotation law and
degrees of total and differential rotation.  They studied the peak
values of GW strain and observed that the \emph{ring-down peak} with
the \emph{largest absolute value} -- which we denote as $H$
hereafter -- is negative ($H \rm < 0$) for models with rapid
differential rotation (and a cylindrical rotation law like our
Eq.~\ref{eq:rotlaw}), while for the rest of their models, $H$ is
positive and coincides with $h_\mathrm{2,pos}$.  They argued that the
detection and extraction of the \emph{sign} of $H$ could
therefore provide clear information about the angular momentum
distribution in the progenitor's core.

Figure~\ref{fig:hpeak2_vs_betaicb} displays $H$ as a function
of $\beta_\mathrm{ic,b}$ for different values of $A$ for all of our
models. ${H}$ grows almost linearly with $\beta_\mathrm{ic,b}$
for $\beta_\mathrm{ic,b} \lesssim 0.08$ for all values of $A$. In this
regime, ${H}$ is positive and corresponds to
$h_\mathrm{2,pos}$. All values of $A$ yield nearly identical $H$
for a given $\beta_\mathrm{ic,b}$ for $\beta_\mathrm{ic,b} \lesssim
0.08$, implying that in this regime $H$ is affected by the total
rotation of the inner core but not by the distribution of angular
momentum within the inner core.

In more rapidly rotating models ($\beta_\mathrm{ic,b} \gtrsim 0.08$),
the values of ${H}$ diverge for different $A$ with the general trend
that \emph{more differentially rotating models yield larger positive}
${H}$. At $\beta_\mathrm{ic,b} \gtrsim 0.12 $, $H$ becomes negative
and no longer corresponds to $h_{2,\mathrm{pos}}$. This occurs first
(in $\beta_\mathrm{ic,b}$) for less differentially rotating models and
the most differentially rotating sequence $A1$ maintains positive $H$
with only a single outlier in which a negative peak has a just
slightly larger magnitude than $h_\mathrm{2,pos}$. From this, we
conclude that the sign of $H$ is \emph{not a good indicator for differential
  rotation}. This is in disagreement with the statement made by
Hayama~\emph{et al.} \cite{hayama:09}, who drew their conclusions on
the basis of a smaller set of models that explored the parameter space
less systematically than our model sequences. In their defense, we
note that $H \rm < 0$ occurs only in models which have at least weak
differential rotation ($A \lesssim A4$ in our model set), simply
because uniformly spinning models cannot reach sufficiently high
$\beta_\mathrm{ic,b}$ for $H$ to become negative
(Figs.~\ref{fig:betaicb_vs_omegac} and
\ref{fig:hpeak2_vs_betaicb}). However, the opposite is not true, since
$H > \rm 0$ does not always indicate uniform rotation.

\begin{figure}[t]
\centering
\includegraphics[width=0.97\columnwidth]{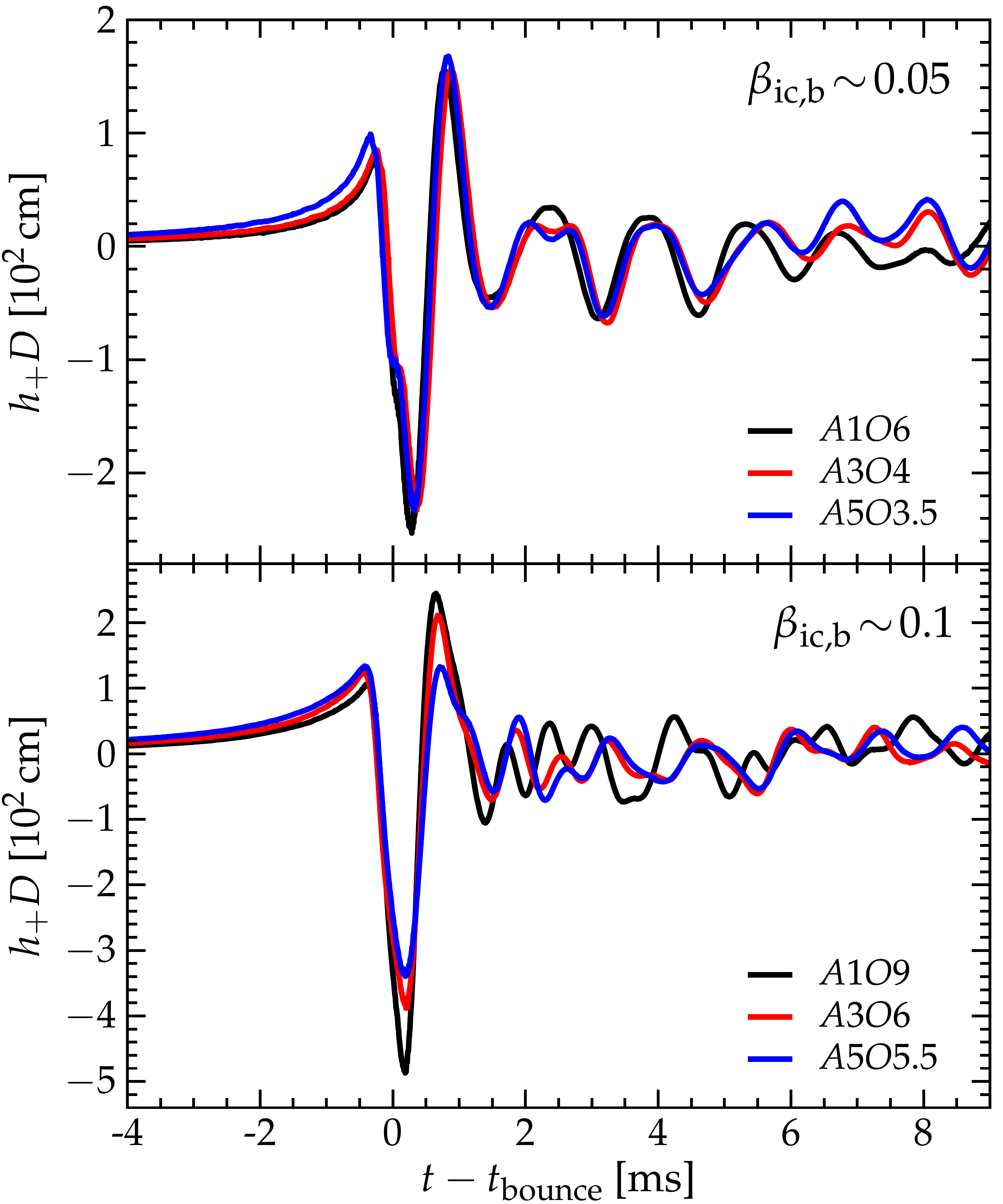}
\caption{GW strain $h_+$ rescaled by source distance $D$.  The top
  panel shows three models with different degrees of differential
  rotation but with the same $\beta_\mathrm{ic,b} \sim 0.05$. The
  bottom panel shows three more rapidly spinning models with
  $\beta_\mathrm{ic,b} \sim 0.1$.  In the first case, the three models
  exhibit almost identical GW signals from bounce, suggesting little
  sensitivity to differential rotation. The situation is different in
  the rapidly rotation case, where there is significant variation 
  between models with different values of the differential rotation
  parameter $A$.}
\label{fig:h_vs_t_slow_vs_rapid}
\vspace{0.5ex}
\end{figure}

\begin{figure}[t]
\centering
\includegraphics[width=0.97\columnwidth]{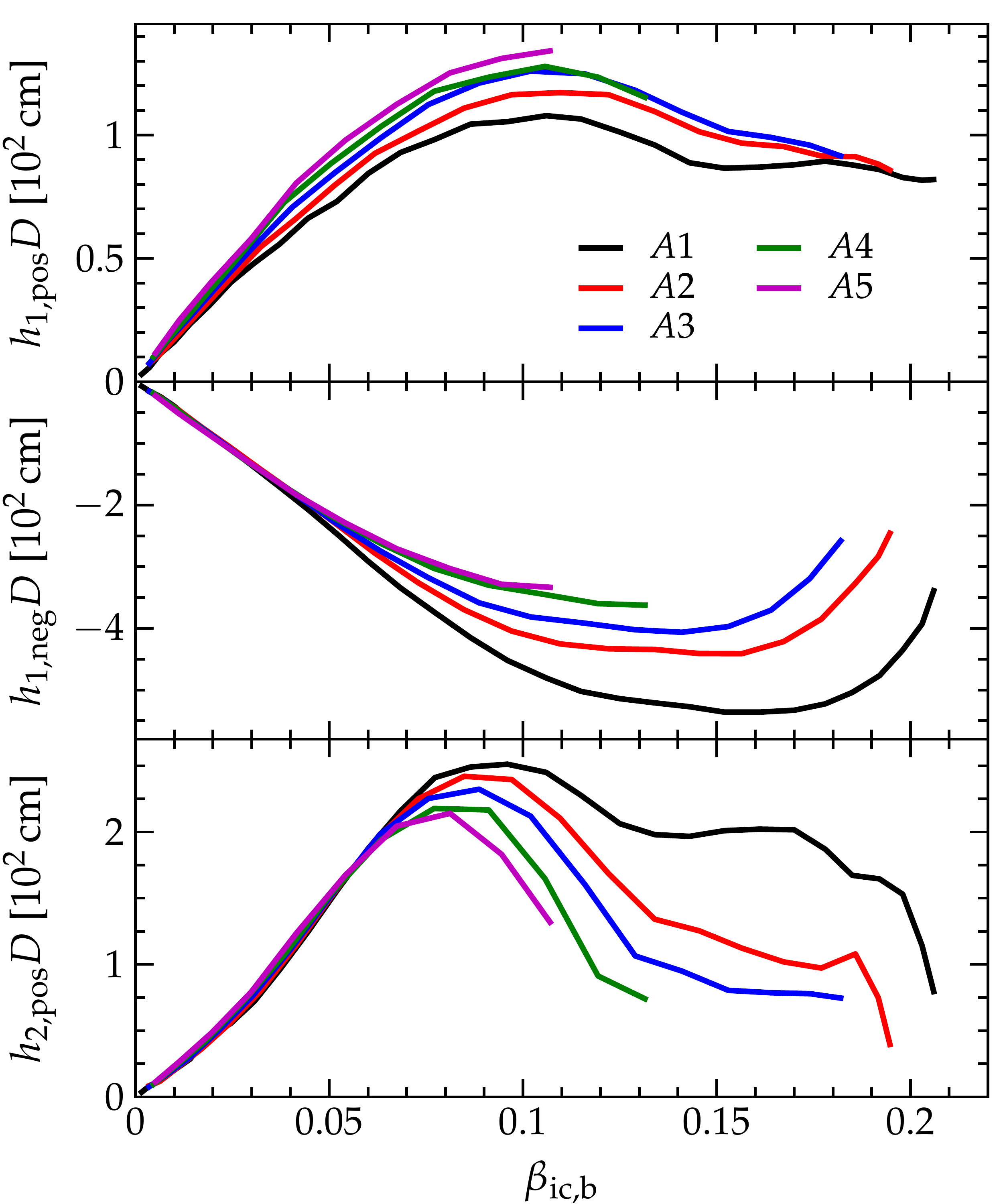}
\caption{The values of the first three peaks of the GW strain
  $h_\mathrm{1,pos}$, $h_\mathrm{1,neg}$, $h_\mathrm{2,pos}$
   (cf.~Fig.~\ref{fig:rho_gwstrain_beta_vs_t}) as a function of 
  $\beta_\mathrm{ic,b}$ plotted for all five model sequences.
  These three prominent GW signal peaks are insensitive to
  the angular momentum distribution for slowly rotating models
  that reach $\beta_\mathrm{ic,b} \lesssim 0.04-0.08$. More rapidly
  rotating models show clear trends with differential rotation.
  }
\label{fig:h3peak_vs_betaicb}
\vspace{0.5ex}
\end{figure}

In Fig.~\ref{fig:h_vs_t_slow_vs_rapid}, we compare waveforms of models
with the same total rotation (as measured by $\beta_\mathrm{ic,b}$)
but different degrees of differential rotation. The top panel depicts
waveforms of models with moderate rotation ($\beta_\mathrm{ic,b} \sim
0.05$) while the bottom panel shows waveforms of rapidly spinning
models with $\beta_\mathrm{ic,b} \sim 0.10$. At $\beta_\mathrm{ic,b}
\sim 0.05$, all choices of $A$ yield essentially the same waveform
between peaks $h_\mathrm{1,pos}$ and $h_\mathrm{2,pos}$ and
differences appear only during the ring-down phase. The situation is
different for rapidly rotating models whose dynamics is more strongly
affected by rotation. While the overall shape of the bounce spike and
its width are still the same for all values of $A$, more differentially
rotating models yield larger $|h_\mathrm{1,neg}|$ and
$h_\mathrm{2,pos}$. The ring-down waveform of the most differentially
rotating model is very different from the other models, reflecting 
the much more pronounced postbounce variations in its central density
shown in Fig.~\ref{fig:rho_vs_t}.

The trends seen for the few select models shown in
Fig.~\ref{fig:h_vs_t_slow_vs_rapid} for the bounce part of the
waveform are very systematic. This is revealed by
Fig.~\ref{fig:h3peak_vs_betaicb}, which shows the values of
$h_\mathrm{1,pos}$, $h_\mathrm{1,neg}$, and $h_\mathrm{2,pos}$ as a
function of $\beta_\mathrm{ic,b}$ for the five considered choices of
differential rotation parameter $A$.  At slow rotation
($\beta_\mathrm{ic,b} \lesssim 0.04-0.08$) there is little dependence
on differential rotation. In more rapidly rotating models, increasing
differential rotation ($=$~decreasing $A$) systematically decreases
$h_\mathrm{1,pos}$, makes $h_\mathrm{1,neg}$ more negative and
increases $h_\mathrm{2,pos}$. This suggests that it should---in
principle---be possible to infer the degree of differential rotation
of rapidly rotating cores from the GW signal alone. In the next
Section~\ref{sec:gwanalysis}, we explore two methods that can be used
to ``measure'' total rotation and $A$ from an observed signal.


\section{Results: Extracting the Angular Momentum Distribution from
an Observed Signal}
\label{sec:gwanalysis}

\subsection{Numerical Template Bank Analysis}
\label{sec:templatebank}

As our analysis in the previous Section suggests, many characteristics
of both the dynamics and GW emission associated with rotating
core-collapse supernovae are dependent on both total rotation
(expressed in $\beta_\mathrm{ic,b}$) and the degree of differential
rotation given by parameter $A$. In the following, we carry out a
\emph{matched filter} analysis to assess the dependence of all signal
features on $\beta_\mathrm{ic,b}$ and $A$ and to study how well we can
hope to extract total and differential rotation from an observed
signal. In the case of a known signal in Gaussian noise, it has been
shown that matched filtering is the optimal detection
technique~\cite{helstrom:68}.  This approach cross-correlates the GW
data observed with a series of filter waveforms, known as
\emph{templates}, produced from GW emission models for the targeted
source.

Generally, GWs from core-collapse supernovae are not amenable to
matched-filtering analysis, since turbulence in the protoneutron star
and behind the stalled shock provides a stochastic component to the
signal \cite{ott:09,kotake:09}. However, in the case of rapid
rotation, convection is suppressed by a stabilizing positive specific
angular momentum gradient in the post-shock region (e.g.,
\cite{ott:08}) and does not contribute significantly to the GW
emission, in particular, not at bounce and in the first few
milliseconds after bounce. Hence, the signal from rotating collapse,
bounce, and postbounce ring-down can be modeled deterministically and
with high precision for a given EOS and neutrino treatment and matched
filtering can be applied.

We construct a \emph{numerical template bank}, utilizing the GW
signals from all models described in Table~\ref{tab:models} (see
Table~\ref{tab:results} for a summary of quantitative results) as
templates to filter observed GW data. Using the known GW waveform
expected from each model and the detector's noise statistics, we find
the best-fitting template for each signal.  We consider signal
waveforms not used as templates in order to imitate the `real-life'
situation where the observed GW signal is not exactly known.  For all
values of $A$, we use injections spanning the template parameter
space, with values of $\Omega_\mathrm{c}$ differing from those of the
templates by at least $0.25\,\mathrm{rad}\,\mathrm{s}^{-1}$.  As
$\beta_\mathrm{ic,b}$ and $A$ for all templates are known, finding the
best-fitting template for an injected signal will infer its associated
closest $\beta_\mathrm{ic,b}$ and $A$. Hereafter, we will refer to
this procedure as ``measuring'' of $\beta_\mathrm{ic,b}$ and $A$.

We perform our analysis in Fourier space, due to frequency dependence
and Gaussian statistics of the GW detector noise, $\tilde{n}$, which
is colored by known one-sided power spectral density (PSD)
$S_{h}(f)$. We model the GW
detector data, $\tilde{d}$, assumed to be comprised of both some
core-collapse supernova GW signal, $\tilde{h}(f;\vec{\lambda})$, and
$\tilde{n}$ as
\begin{equation}
\label{eq:di}
\tilde{d_{i}} = \tilde{h}(f_{i};\vec{\lambda}) + \tilde{n}_{i}\,\,,
\end{equation}
where $i$ denotes the frequency bin index.

The parameter dependence of the GW signals considered here is encoded in
$\vec{\lambda}$,
\begin{equation}
\label{eq:theta}
\vec{\lambda} =
\{D,t_{0},\iota,\xi,\theta,\phi,\psi\},
\end{equation}
where $D$ is the source distance, $t_{0}$ is the time at which the GW
signal arrives at the detector, and $(\iota,\xi,\theta,\phi,\psi)$ are
source angles. Here, $(\iota,\xi)$ relate the preferred internal axes
of the source to the location of the detector, $(\theta,\phi)$ relate
the preferred internal axes of the detector to the location of the
source and $\psi$ defines the relationship between the source and the
detector, via the plane characterizing the polarization of emitted
GWs~\cite{mtw}.


\begin{longtable*}{lccccccccccccc}
  \caption{Summary of simulation results. $\Omega_\mathrm{c}$ is the
    intitial central angular velocity, $\rho_\mathrm{c,b}$ is the
    central density at bounce, $\rho_\mathrm{c,pb}$ is the early
    postbounce central density, $\rho_\mathrm{max,pb}$ is the
    postbounce maximum density, $\beta_\mathrm{ic,b}$ and
    $\beta_\mathrm{ic,pb}$ are ratios of the rotational kinetic
    energy to the gravitational binding energy of the inner core at
    bounce and early postbounce phase, respectively. $M_\mathrm{ic,b}$
    and $J_\mathrm{ic,b}$ are the inner core mass and angular momentum
    at bounce, $|h_{+,2}|D$ is the second peak of the GW signal, while
    $|h_\mathrm{+,max}|D$ is its maximum value. $f_\mathrm{max}$ is the
    frequency at which the GW spectral energy density reaches a
    maximum value. The symbol ${}^*$ at the end of the model name
    indicates that for this model the peak GW signal is produced by
    convection.} \label{tab:results}\\   
  \hline\hline
  Model&$\Omega_\mathrm{c}$&$\rho_\mathrm{c,b}$&$\rho_\mathrm{c,pb}$&$\rho_\mathrm{max,pb}$&$\beta_\mathrm{ic,b}$&$\beta_\mathrm{ic,pb}$&$M_\mathrm{ic,b}$&$J_\mathrm{ic,b}$&${H}D$&$|h_\mathrm{+,max}|D$&$E_\mathrm{GW}$&$f_\mathrm{max}$\\
       &[$\mathrm{rad\,s^{-1}}$]&[$10^{14}$          &[$10^{14}$            &[$10^{14}$            &[$10^{-2}$]&[$10^{-2}$]&[$M_\odot$]&[$10^{48}$        &[cm]&[cm]&[$10^{-9}$    &[Hz]\\
       &                        &$\mathrm{gcm^{-3}}$]&$\mathrm{g\,cm^{-3}}$]&$\mathrm{g\,cm^{-3}}$]&           &           &           &$\mathrm{erg\,s}$]&    &    &$M_\odot c^2$]&    \\
  \hline
  \endfirsthead
  \caption{Continued}\\
  \hline\hline
  Model&$\Omega_\mathrm{c}$&$\rho_\mathrm{c,b}$&$\rho_\mathrm{c,pb}$&$\rho_\mathrm{max,pb}$&$\beta_\mathrm{ic,b}$&$\beta_\mathrm{ic,pb}$&$M_\mathrm{ic,b}$&$J_\mathrm{ic,b}$&${H}D$&$|h_\mathrm{+,max}|D$&$E_\mathrm{GW}$&$f_\mathrm{max}$\\
       &[$\mathrm{rad\,s^{-1}}$]&[$10^{14}$          &[$10^{14}$            &[$10^{14}$            &[$10^{-2}$]&[$10^{-2}$]&[$M_\odot$]&[$10^{48}$        &[cm]&[cm]&[$10^{-9}$    &[Hz]\\
       &                        &$\mathrm{gcm^{-3}}$]&$\mathrm{g\,cm^{-3}}$]&$\mathrm{g\,cm^{-3}}$]&           &           &           &$\mathrm{erg\,s}$]&    &    &$M_\odot c^2$]&    \\
  \hline
  \endhead
  \hline
  \hline
  \multicolumn{11}{r}{Continued on Next page...}
  \endfoot
  \hline
  \hline
  \endlastfoot

$A1O1^*  $& 1.0  & 4.39 & 3.60 & 3.60 &  0.16  &  0.13  & 0.58 &  0.31 & 44.44  &\z7.38   &  7.19  & 829.17 \\
$A1O1.5^*$& 1.5  & 4.38 & 3.59 & 3.59 &  0.36  &  0.30  & 0.58 &  0.46 & 44.53  & 15.30   &  8.62  & 821.17 \\
$A1O2^*  $& 2.0  & 4.35 & 3.57 & 3.57 &  0.64  &  0.53  & 0.59 &  0.63 & 44.65  & 24.29   &  9.02  & 937.69 \\
$A1O2.5^*$& 2.5  & 4.36 & 3.55 & 3.55 &  1.00  &  0.82  & 0.58 &  0.77 & 44.81  & 39.67   &  6.26  & 817.92 \\
$A1O3^*  $& 3.0  & 4.35 & 3.52 & 3.52 &  1.41  &  1.17  & 0.58 &  0.92 & 45.00  & 61.77   &  7.86  & 842.68 \\
$A1O3.5  $& 3.5  & 4.32 & 3.49 & 3.49 &  1.90  &  1.58  & 0.59 &  1.11 & 45.23  & 83.27   &  6.75  & 824.79 \\
$A1O4    $& 4.0  & 4.26 & 3.46 & 3.46 &  2.46  &  2.04  & 0.61 &  1.37 & 45.49  & 109.77\z&  8.90  & 764.99 \\
$A1O4.5  $& 4.5  & 4.22 & 3.42 & 3.42 &  3.07  &  2.54  & 0.61 &  1.50 & 45.80  & 138.94\z& 12.01\z& 833.67 \\
$A1O5    $& 5.0  & 4.20 & 3.39 & 3.39 &  3.73  &  3.08  & 0.61 &  1.67 & 46.15  & 171.21\z& 12.60\z& 678.00 \\
$A1O5.5  $& 5.5  & 4.15 & 3.34 & 3.35 &  4.45  &  3.66  & 0.65 &  1.99 & 46.54  & 207.07\z& 19.05\z& 681.96 \\
$A1O6    $& 6.0  & 4.08 & 3.29 & 3.29 &  5.20  &  4.27  & 0.65 &  2.17 & 46.98  & 246.82\z& 31.39\z& 716.50 \\
$A1O6.5  $& 6.5  & 4.03 & 3.24 & 3.25 &  6.01  &  4.92  & 0.65 &  2.39 & 47.47  & 291.49\z& 38.88\z& 816.62 \\
$A1O7    $& 7.0  & 4.00 & 3.17 & 3.18 &  6.84  &  5.58  & 0.67 &  2.69 & 48.01  & 334.38\z& 44.36\z& 764.27 \\
$A1O7.5  $& 7.5  & 3.92 & 3.11 & 3.12 &  7.73  &  6.28  & 0.68 &  2.94 & 48.60  & 374.54\z& 59.05\z& 786.14 \\
$A1O8    $& 8.0  & 3.85 & 3.04 & 3.06 &  8.65  &  6.98  & 0.70 &  3.24 & 49.26  & 415.24\z& 66.28\z& 811.45 \\
$A1O8.5  $& 8.5  & 3.74 & 2.97 & 3.00 &  9.60  &  7.70  & 0.70 &  3.50 & 49.98  & 452.12\z& 73.11\z& 834.94 \\
$A1O9    $& 9.0  & 3.65 & 2.89 & 2.93 & 10.60\z&  8.42  & 0.72 &  3.87 & 50.77  & 480.53\z& 80.65\z& 844.19 \\
$A1O9.5  $& 9.5  & 3.56 & 2.81 & 2.85 & 11.50\z&  9.14  & 0.74 &  4.17 & 51.64  & 502.39\z& 86.60\z& 827.66 \\
$A1O10   $&10.0\z& 3.45 & 2.71 & 2.77 & 12.50\z&  9.85  & 0.74 &  4.52 & 52.60  & 514.13\z& 87.51\z& 562.73 \\
$A1O10.5 $&10.5\z& 3.35 & 2.62 & 2.69 & 13.40\z& 10.57\z& 0.76 &  4.92 & 53.66  & 520.96\z& 86.21\z& 560.75 \\
$A1O11   $&11.0\z& 3.23 & 2.53 & 2.61 & 14.30\z& 11.28\z& 0.78 &  5.28 & 54.82  & 527.25\z& 78.28\z& 504.13 \\
$A1O11.5 $&11.5\z& 3.14 & 2.48 & 2.55 & 15.20\z& 12.02\z& 0.79 &  5.71 & 56.13  & 535.99\z& 76.40\z& 483.85 \\
$A1O12   $&12.0\z& 3.04 & 2.46 & 2.50 & 16.10\z& 12.78\z& 0.80 &  6.15 & 57.58  & 535.99\z& 74.46\z& 483.08 \\
$A1O12.5 $&12.5\z& 3.00 & 2.41 & 2.44 & 17.00\z& 13.57\z& 0.82 &  6.63 & 59.21  & 532.98\z& 70.81\z& 477.07 \\
$A1O13   $&13.0\z& 2.91 & 2.34 & 2.37 & 17.80\z& 14.30\z& 0.84 &  7.10 & 61.02  & 522.60\z& 62.99\z& 448.28 \\
$A1O13.5 $&13.5\z& 2.82 & 2.25 & 2.28 & 18.50\z& 14.96\z& 0.85 &  7.58 &-63.06\p& 504.03\z& 50.41\z& 433.69 \\
$A1O13   $&14.0\z& 2.72 & 2.15 & 2.18 & 19.20\z& 15.55\z& 0.86 &  8.06 & 65.37  & 476.98\z& 37.97\z& 387.50 \\
$A1O14.5 $&14.5\z& 2.64 & 2.05 & 2.08 & 19.80\z& 16.10\z& 0.89 &  8.69 & 68.01  & 435.46\z& 25.70\z& 375.43 \\
$A1O15   $&15.0\z& 2.53 & 1.89 & 1.92 & 20.30\z& 16.41\z& 0.91 &  9.33 & 71.01  & 393.11\z& 17.24\z& 319.46 \\
$A1O15.5 $&15.5\z& 2.41 & 1.69 & 1.72 & 20.60\z& 16.47\z& 0.94 & 10.24\z & 74.53& 339.02\z& 11.41\z& 271.36 \\
\hline                                                                                    
$A2O1^*  $& 1.0 & 4.42 & 3.59 & 3.59 &  0.36  &  0.31  & 0.57 & 0.44 &\z0.98  & 16.12   & 17.36\z& 851.53 \\
$A2O1.5^*$& 1.5 & 4.42 & 3.57 & 3.57 &  0.63  &  0.54  & 0.58 & 0.61 &\z1.71  & 24.91   & 16.34\z& 859.90 \\
$A2O2^*  $& 2.0 & 4.31 & 3.54 & 3.55 &  1.10  &  0.95  & 0.58 & 0.81 &\z3.01  & 45.35   &  7.59  & 837.97 \\
$A2O2.5^*$& 2.5 & 4.28 & 3.51 & 3.51 &  1.70  &  1.46  & 0.59 & 1.04 &\z4.65  & 73.49   & 16.82\z& 777.70 \\
$A2O3    $& 3.0 & 4.26 & 3.48 & 3.48 &  2.42  &  2.06  & 0.61 & 1.31 &\z6.60  & 104.90\z&  8.57  & 853.71 \\
$A2O3.5  $& 3.5 & 4.19 & 3.43 & 3.43 &  3.23  &  2.74  & 0.61 & 1.51 &\z8.83  & 142.33\z& 11.67\z& 745.77 \\
$A2O4    $& 4.0 & 4.14 & 3.38 & 3.39 &  4.14  &  3.49  & 0.63 & 1.86 & 11.32  & 182.76\z& 14.81\z& 687.60 \\
$A2O4.5  $& 4.5 & 4.07 & 3.33 & 3.33 &  5.13  &  4.31  & 0.64 & 2.12 & 14.01  & 228.38\z& 26.08\z& 827.37 \\
$A2O5    $& 5.0 & 4.00 & 3.26 & 3.26 &  6.18  &  5.18  & 0.65 & 2.46 & 16.89  & 278.38\z& 34.18\z& 696.93 \\
$A2O5.5  $& 5.5 & 3.91 & 3.19 & 3.19 &  7.30  &  6.12  & 0.67 & 2.85 & 19.95  & 326.18\z& 45.00\z& 732.25 \\
$A2O6    $& 6.0 & 3.80 & 3.11 & 3.12 &  8.48  &  7.08  & 0.70 & 3.25 & 23.17  & 369.89\z& 54.82\z& 763.68 \\
$A2O6.5  $& 6.5 & 3.69 & 3.03 & 3.03 &  9.71  &  8.06  & 0.72 & 3.72 & 26.53  & 404.59\z& 58.80\z& 782.64 \\
$A2O7    $& 7.0 & 3.58 & 2.93 & 2.94 & 10.96\z&  9.03  & 0.72 & 4.09 & 29.94  & 425.35\z& 60.28\z& 797.74 \\
$A2O7.5  $& 7.5 & 3.45 & 2.84 & 2.86 & 12.21\z& 10.02\z& 0.74 & 4.52 & 33.36  & 433.27\z& 59.28\z& 814.41 \\
$A2O8    $& 8.0 & 3.30 & 2.73 & 2.75 & 13.40\z& 10.95\z& 0.76 & 5.00 &-36.61\p& 434.36\z& 53.48\z& 820.71 \\
$A2O8.5  $& 8.5 & 3.17 & 2.63 & 2.65 & 14.55\z& 11.88\z& 0.78 & 5.63 &-39.75\p& 440.92\z& 47.17\z& 828.62 \\
$A2O9    $& 9.0 & 3.04 & 2.52 & 2.54 & 15.65\z& 12.80\z& 0.80 & 6.35 &-42.75\p& 441.19\z& 40.59\z& 830.37 \\
$A2O9.5  $& 9.5 & 2.88 & 2.44 & 2.45 & 16.73\z& 13.80\z& 0.82 & 6.90 &-45.70\p& 421.52\z& 32.56\z& 394.97 \\
$A2O10   $&10.0 & 2.77 & 2.33 & 2.34 & 17.70\z& 14.70\z& 0.84 & 7.62 &-48.35\p& 385.19\z& 21.90\z& 373.13 \\
$A2O10.5 $&10.5 & 2.65 & 2.19 & 2.20 & 18.58\z& 15.55\z& 0.86 & 8.37 & 50.76  & 327.00\z& 14.20\z& 330.47 \\
$A2O11   $&11.0 & 2.52 & 2.02 & 2.03 & 19.17\z& 16.21\z& 0.87 & 9.05 &-52.37\p& 283.57\z&  8.98  & 292.98 \\
$A2O11.5 $&11.5 & 2.33 & 1.75 & 1.76 & 19.47\z& 16.38\z& 0.89 & 9.76 &-53.19\p& 245.87\z&  5.39  & 216.11 \\
\hline                                                                        
$A3O1^*  $ & 1.0 & 4.47 & 3.59 & 3.59 &  0.36  &  0.31  & 0.57 & 0.45 & 44.59    & 15.84   &  7.05  & 873.82 \\
$A3O1.5^*$ & 1.5 & 4.38 & 3.56 & 3.57 &  0.80  &  0.70  & 0.58 & 0.69 & 44.87    & 32.13   &  8.62  & 850.91 \\
$A3O2^*  $ & 2.0 & 4.26 & 3.53 & 3.53 &  1.40  &  1.23  & 0.59 & 0.94 & 45.27    & 60.87   &  4.04  & 798.47 \\
$A3O2.5^*$ & 2.5 & 4.27 & 3.49 & 3.50 &  2.15  &  1.88  & 0.61 & 1.24 & 45.79    & 94.90   &  6.50  & 772.97 \\
$A3O3    $ & 3.0 & 4.15 & 3.45 & 3.45 &  3.03  &  2.63  & 0.60 & 1.45 & 46.44    & 135.69\z& 16.04\z& 873.48 \\
$A3O3.5  $ & 3.5 & 4.12 & 3.40 & 3.40 &  4.04  &  3.48  & 0.63 & 1.82 & 47.26    & 178.53\z& 14.12\z& 718.34 \\
$A3O4    $ & 4.0 & 4.04 & 3.33 & 3.33 &  5.14  &  4.41  & 0.65 & 2.18 & 48.25    & 227.84\z& 23.05\z& 706.91 \\
$A3O4.5  $ & 4.5 & 3.96 & 3.27 & 3.27 &  6.32  &  5.42  & 0.66 & 2.55 & 49.44    & 274.55\z& 34.01\z& 706.90 \\
$A3O5    $ & 5.0 & 3.85 & 3.18 & 3.18 &  7.56  &  6.48  & 0.68 & 2.98 & 50.88    & 317.99\z& 44.21\z& 725.72 \\
$A3O5.5  $ & 5.5 & 3.74 & 3.09 & 3.10 &  8.87  &  7.59  & 0.70 & 3.40 & 52.60    & 358.69\z& 49.45\z& 749.09 \\
$A3O6    $ & 6.0 & 3.62 & 3.00 & 3.00 & 10.20\z&  8.70  & 0.71 & 3.81 & 54.70    & 381.64\z& 49.92\z& 767.59 \\
$A3O6.5  $ & 6.5 & 3.49 & 2.89 & 2.90 & 11.60\z&  9.80  & 0.73 & 4.40 & 57.28    & 391.75\z& 48.03\z& 780.76 \\
$A3O7    $ & 7.0 & 3.34 & 2.79 & 2.80 & 12.90\z& 10.88\z& 0.75 & 4.92 &-60.51\p  & 402.40\z& 43.74\z& 795.01 \\
$A3O7.5  $ & 7.5 & 3.18 & 2.66 & 2.67 & 14.10\z& 11.86\z& 0.77 & 5.60 &-64.66\p  & 406.50\z& 36.92\z& 799.64 \\
$A3O8    $ & 8.0 & 3.04 & 2.55 & 2.56 & 15.30\z& 12.90\z& 0.79 & 6.31 &-70.24\p  & 397.21\z& 31.53\z& 797.59 \\
$A3O8.5  $ & 8.5 & 2.90 & 2.44 & 2.45 & 16.40\z& 13.96\z& 0.81 & 7.02 &-78.19\p  & 370.71\z& 22.83\z& 792.28 \\
$A3O9    $ & 9.0 & 2.75 & 2.31 & 2.32 & 17.40\z& 14.94\z& 0.83 & 7.74 &-90.70\p  & 319.63\z& 13.36\z& 371.50 \\
$A3O9.5  $ & 9.5 & 2.57 & 2.16 & 2.16 & 18.20\z& 15.87\z& 0.85 & 8.61 &-112.97\pz& 258.16\z&  7.88  & 285.59 \\
\hline
$A4O1^*  $& 1.0 & 4.36 & 3.58 & 3.58 &  0.47  &  0.42  & 0.57 & 0.51 &  1.27   & 18.49   &  4.18  & 871.56 \\
$A4O1.5^*$& 1.5 & 4.31 & 3.55 & 3.55 &  1.03  &  0.95  & 0.58 & 0.78 &  2.82   & 43.16   &  3.66  & 971.44 \\
$A4O2^*  $& 2.0 & 4.26 & 3.51 & 3.51 &  1.80  &  1.65  & 0.60 & 1.10 &  4.91   & 80.59   &  6.42  & 808.24 \\
$A4O2.5^*$& 2.5 & 4.16 & 3.46 & 3.46 &  2.74  &  2.48  & 0.60 & 1.35 &  7.49   & 123.48\z&  9.11  & 836.50 \\
$A4O3    $& 3.0 & 4.11 & 3.41 & 3.41 &  3.84  &  3.43  & 0.63 & 1.75 & 10.49\z & 169.65\z& 11.64\z& 731.20 \\
$A4O3.5  $& 3.5 & 4.03 & 3.33 & 3.33 &  5.05  &  4.49  & 0.64 & 2.12 & 13.80\z & 218.27\z& 20.80\z& 700.07 \\
$A4O4    $& 4.0 & 3.92 & 3.26 & 3.26 &  6.34  &  5.65  & 0.66 & 2.53 & 17.33\z & 262.80\z& 31.50\z& 696.11 \\
$A4O4.5  $& 4.5 & 3.80 & 3.17 & 3.17 &  7.71  &  6.86  & 0.67 & 2.90 & 21.05\z & 302.96\z& 39.80\z& 717.07 \\
$A4O5    $& 5.0 & 3.69 & 3.06 & 3.07 &  9.12  &  8.07  & 0.70 & 3.46 & 24.91\z & 330.28\z& 41.04\z& 727.72 \\
$A4O5.5  $& 5.5 & 3.54 & 2.97 & 2.97 & 10.57\z&  9.30  & 0.72 & 4.05 & 28.88\z & 345.31\z& 38.68\z& 748.63 \\
$A4O6    $& 6.0 & 3.39 & 2.86 & 2.86 & 11.94\z& 10.47\z& 0.73 & 4.56 &-32.62\pz& 360.06\z& 35.27\z& 749.91 \\
$A4O6.5  $& 6.5 & 3.26 & 2.74 & 2.74 & 13.15\z& 11.51\z& 0.75 & 5.20 &-35.92\pz& 362.52\z& 31.14\z& 759.49 \\
\hline                                                                                 
$A5O1^*  $& 1.0 & 4.36 & 3.58 & 3.58 & 0.52  & 0.48 & 0.58 & 0.54 & 44.92\z  & 22.67   &  3.85  & 299.25 \\
$A5O1.5^*$& 1.5 & 4.31 & 3.54 & 3.55 & 1.13  & 1.07 & 0.58 & 0.81 & 45.64\z  & 52.18   &  4.01  & 855.35 \\
$A5O2^*  $& 2.0 & 4.20 & 3.50 & 3.50 & 1.97  & 1.84 & 0.60 & 1.14 & 46.71\z  & 88.98   &  5.42  & 850.52 \\
$A5O2.5  $& 2.5 & 4.16 & 3.45 & 3.45 & 2.99  & 2.75 & 0.61 & 1.42 & 48.21\z  & 133.40\z&  7.67  & 716.12 \\
$A5O3    $& 3.0 & 4.08 & 3.39 & 3.39 & 4.15  & 3.77 & 0.63 & 1.57 & 50.29\z  & 183.31\z& 13.83\z& 703.91 \\
$A5O3.5  $& 3.5 & 3.99 & 3.31 & 3.31 & 5.42  & 4.90 & 0.64 & 2.18 & 53.19\z  & 229.20\z& 23.55\z& 701.79 \\
$A5O4    $& 4.0 & 3.87 & 3.24 & 3.24 & 6.74  & 6.10 & 0.67 & 2.67 & 57.43\z  & 270.92\z& 33.99\z& 712.85 \\
$A5O4.5  $& 4.5 & 3.74 & 3.15 & 3.15 & 8.12  & 7.32 & 0.68 & 3.12 & 64.16\z  & 303.24\z& 39.64\z& 721.69 \\
$A5O5    $& 5.0 & 3.64 & 3.06 & 3.06 & 9.45  & 8.49 & 0.70 & 3.61 & 77.44\z  & 328.64\z& 37.84\z& 739.29 \\
$A5O5.5  $& 5.5 & 3.53 & 2.96 & 2.96 &10.70\z& 9.54 & 0.71 & 4.05 & 161.22\zz& 333.56\z& 35.93\z& 760.41   
\end{longtable*}

Our goal is to establish the best-fitting template for the observed GW data.
We construct the noise-weighted inner product, $\langle d,x^{j} \rangle$, 
for all templates $\tilde{x}^{j}(f)$ with the data, where $j$ denotes
template index in the catalog, as
\begin{align}
\langle d,x^{j} \rangle &=
2 \max_{t_{0}} \int_{-\infty}^{\infty} \frac{\tilde{d}(f)\tilde{x}^{j}(f)^{*}e^{i2\pi
ft_{0}}}{S_{h}(f)}\mathrm{d}f\,\,,
\end{align}
where ${}^{*}$ denotes complex conjugation.  We assume stationary, Gaussian 
detector noise. We
numerically maximize this quantity over all possible $t_{0}$ using fast-Fourier
transforms.  From this, we 
compute the detection \emph{signal-to-noise ratio} 
(SNR) for each template, $\rho^{j}$, as
\begin{align}
\rho^{j} &= \frac{\langle d,x^{j} \rangle}{\langle x^{j},x^{j}
\rangle^{1/2}}\,\,,
\end{align}
where $\langle x^{j},x^{j} \rangle$ is the template norm.  For the
optimal case in which $h = x^{j}$, the expected signal SNR is simply
$\langle x^{j},x^{j} \rangle^{1/2}$.  Given this quantity, we
calculate the set of $\rho^{j}$ across all templates to determine the
best-fitting template, given the data observed.  We define the best-fitting
template as the template $j$ producing the largest $\rho^{j}$, given an imposed
detection threshold of $\rho_{j} \geq 8$~\cite{gonzalez:04}.

We utilize simulated Gaussian noise colored by the zero-detuned high
power configuration of aLIGO~\cite{LIGO-sens-2010}, and, for
simplicity, consider a single GW detector.  We repeat all calculations
with ten different realizations of detector noise and report the
averaged result. We assume that the source is optimally-oriented, and
located relative to the detector such that the observed GW strain,
$h$, given by
\begin{align}
h &= h_{+}F_{+} + h_{\times}F_{\times}\,\,,
\end{align}
is maximized, where the antenna response functions, $F_{+}$ and $F_{\times}$, are 
given by
\begin{align}
F_{+} &= \frac{1}{2}(1 + \cos^{2}\theta)\cos 2\phi\cos 2 \psi -
\cos\theta\sin2\phi\sin 2\psi,\,\,\\ 
F_{\times} &= \frac{1}{2}(1 + \cos^{2}\theta)\cos 2\phi\sin 2 \psi +
\cos\theta\sin 2\phi\cos 2\psi,\,\,
\end{align}
respectively, and $h_{+}$ is related to $H_{20}$, the $(l,m)=(2,0)$ mode of the
GW multipole expansion~\cite{ajith:07}, as
\begin{align}
h_{+} &= \frac{1}{D}\sqrt{\frac{15}{32\pi}}H_{20}\sin^{2}\iota\,\,.
\end{align}
Due to the axisymmetric nature of the simulations presented here,
$h_{+}$ is independent of $\xi$ and all GW emission will be linearly
polarized (i.e. $h_{\times} = 0$).  In physical terms, these
assumptions correspond to setting source angles
$(\iota,\theta,\phi,\psi) = (\pi/2,0,0,0)$.  We place all sources at a
known distance  $D=10\,\mathrm{kpc}$, restricting our analysis to the
galactic locus.

To conclude the discussion of our numerical template bank analysis, we
note that the nature of our analysis is fundamentally distinct from
template banks used in the context of LIGO/Virgo GW searches for
compact binary coalescences, which can produce templates
``on-the-fly'' for binary inspiral signals using post-Newtonian
expressions for the GW strain for arbitrary system
parameters~\cite{cutler:93}.  The GW emission from rotating core
collapse is complicated, dependent on many parameters, and has yet to
be described phenomenologically.  This means that the span of the
numerical template bank across the simulation parameter space is
limited to discrete samples, with template waveforms produced by
simulations of core collapse.  The nature of templates for binary
inspirals also conveys that the phase of GW emission can be robustly
predicted, whereas convection in the later postbounce stages of core
collapse is largely stochastic, resulting in unpredictable waveform
phase.  This limits the predictive power of our analysis in slowly
rotating models in which convection is abound. Additionally, the study
presented here considers only two unknown progenitor parameters ($A$
and $\beta_{\mathrm{ic,b}}$), while in reality, the simulation
parameter space is larger and also includes (but is not necessarily
limited to) EOS and electron fraction parameterization.

\subsubsection{Extraction of $\beta_\mathrm{ic,b}$}

\begin{figure}[t]
\centering
\includegraphics[width=0.97\columnwidth]{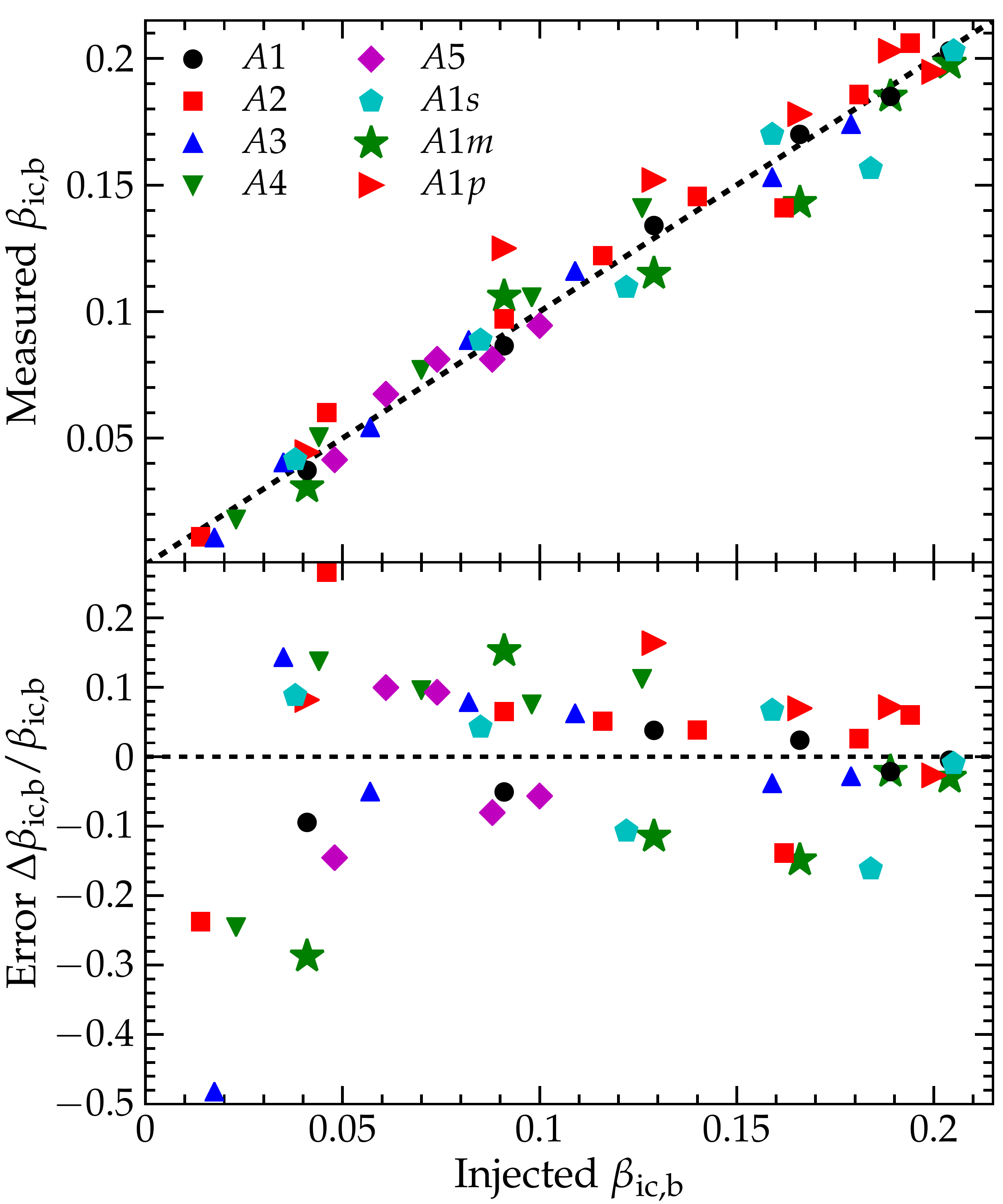}
\caption{Upper panel: Measured $\beta_\mathrm{ic,b}$ as a function of
  true $\beta_\mathrm{ic,b}$ for all injected waveforms. The dashed
  black line denotes the optimal case in which the measured and true
  $\beta_\mathrm{ic,b}$ are equal.  Lower panel: The relative
  deviation of $\beta_\mathrm{ic,b}$ measured from its true value.
  For most signals, we find that $\beta_\mathrm{ic,b}$ is measured
  with $\sim 10-20\%$ accuracy. The errors are largest for slowly
  rotating models since these have strong stochastic convective
  components in their waveforms. Outliers at more rapid rotation are
  signals from the $A1s$, $A1m$ and $A1p$ injection sets.  The $A1$s
  model uses the Shen~\emph{et~al.}  EOS~\cite{shen:98a,shen:98b}
  rather than the Lattimer-Swesty EOS~\cite{lseos:91} used for the
  fiducial models, while the $A1$m and $A1$p models are simulated with
  $\sim 5\,\%$ decreased and increased $Y_\mathrm{e}(\rho)$,
  respectively, at nuclear densities.}
\label{fig:betaicbmeasuredMF}
\vspace{0.5ex}
\end{figure}

\begin{figure}[t]
\centering
\includegraphics[width=0.97\columnwidth]{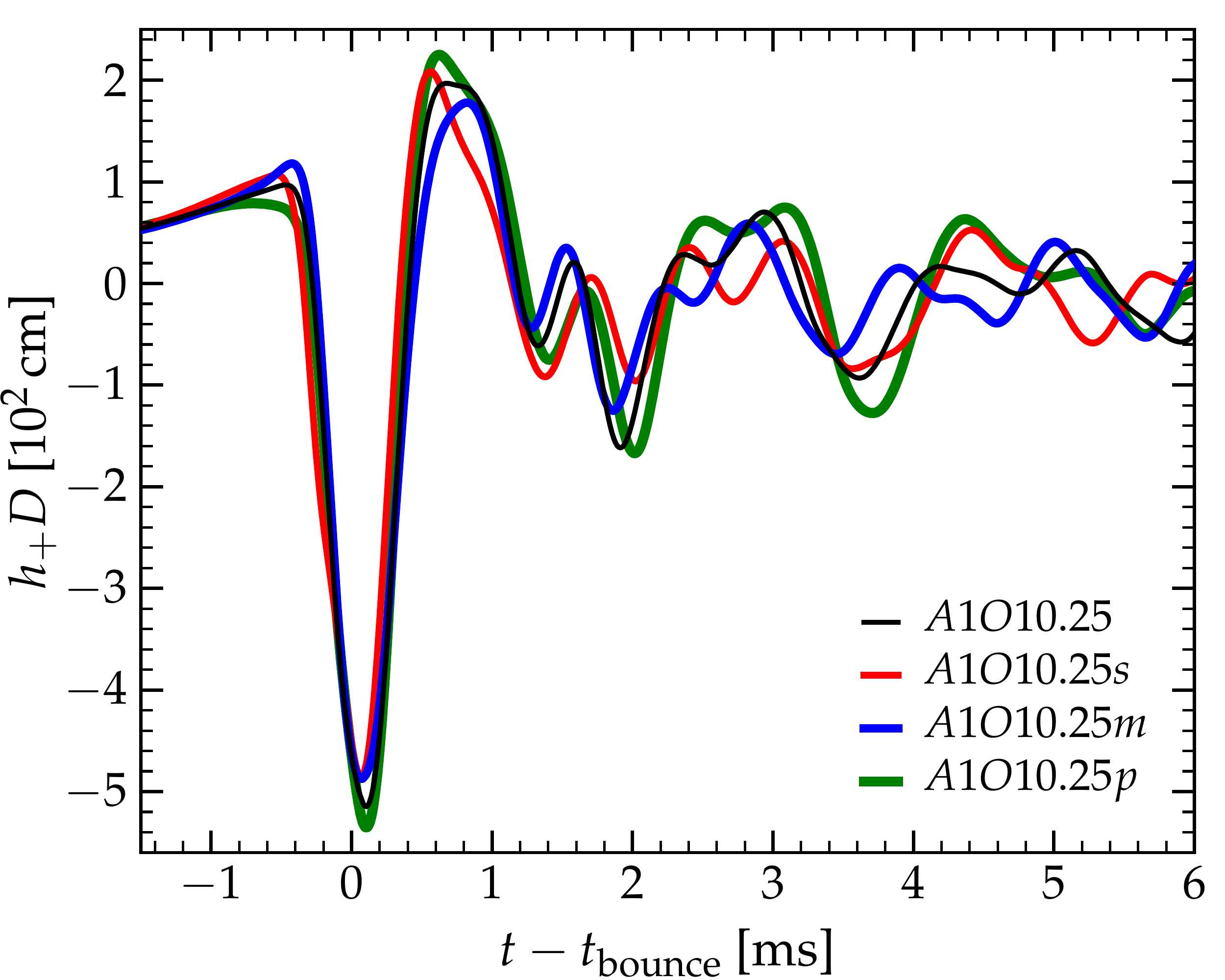}
\caption{GW strain $h_{+}$ rescaled by source distance $D$ for
  injected waveforms $A1O10.25$, $A1O10.25s$, $A1O10.25m$ and
  $A1O10.25p$. The black line represents the waveform generated using
  the Lattimer-Swesty EOS~\cite{lseos:91} and the standard
  $Y_\mathrm{e}(\rho)$ parametrization, the red graph corresponds to
  the model simulated with the Shen~\emph{et~al.}
  EOS~\cite{shen:98a,shen:98b}, while the blue and green graphs are
  simulated with $\sim 5\,\%$ increased and decreased
  $Y_\mathrm{e}(\rho)$, respectively, at nuclear densities.}
\label{fig:h_vs_t_A1O10.25}
\vspace{0.5ex}
\end{figure}

The upper panel of Fig.~\ref{fig:betaicbmeasuredMF} shows
$\beta_\mathrm{ic,b}$ measured for injected waveforms versus the true
values of $\beta_\mathrm{ic,b}$ of those models.  The dashed black
line denotes the optimal case in which the measured and true
$\beta_\mathrm{ic,b}$ are identical.  The lower panel shows the
relative deviation of $\beta_\mathrm{ic,b}$ from its correct
value. For most injected waveforms, the value of $\beta_\mathrm{ic,b}$
measured lies within $\sim 20\,\%$ of the true value for the five
values of $A$ considered, with $\beta_\mathrm{ic,b}$ ranging from
$\sim 0.01$ to $\sim 0.2$.  The average relative deviation of measured
$\beta_\mathrm{ic,b}$ from its true value is $\sim 8\,\%$ for all
injected waveforms.  The measurement error is
  largest in slowly spinning models (small $\beta_\mathrm{ic,b}$),
  because these emit GW signals with strong stochastic components from
  prompt postbounce convection.

The matched filter analysis can extract $\beta_\mathrm{ic,b}$ with
good accuracy across a wide range of both total rotation and
differential rotation.  This is not surprising, since we showed in
Section \ref{sec:gw_peaks} that the GW signal amplitudes depend
primarily on $\beta_\mathrm{ic,b}$ both for slowly and rapidly
rotating models.

To test the robustness of this conclusion, we explore the accuracy
with which this analysis can extract $\beta_\mathrm{ic,b}$ for
injected signals that are produced using a different nuclear EOS or
different $Y_\mathrm{e}(\rho)$
parametrization.  Differences in these aspects
  are associated with differences in the pressure, energy density, and
  other thermodynamic quantities. This leads to variations in the 
  mass of the inner core at bounce ($M_{\mathrm{ic,b}}$) and
  influences the dynamics of the final phase of collapse, bounce, and
  ring-down oscillations. The EOS dependence of GW emission from
  rotating core collapse was first explored by~\cite{dimmelmeier:08},
  while the influence of the $Y_\mathrm{e}$ parametrization was studied in
  the context of accretion-induced collapse (AIC) by
  \cite{abdikamalov:10}.

To evaluate the dependence of our results on the EOS, we reproduce
signals for injection from the $A1$ model sequence using the
Shen~\emph{et~al.}~\cite{shen:98a,shen:98b} EOS in place of the
Lattimer \& Swesty EOS~\cite{lseos:91} used for the fiducial models
listed in Tables~\ref{tab:models}-\ref{tab:results}, while keeping the
$Y_\mathrm{e}(\rho)$ parametrization unchanged.  We hereafter refer to
this set of injections as $A1s$.  To explore the dependence of the GW
signals on the $Y_\mathrm{e}$ parametrization, we repeat the same
sequence with the Lattimer \& Swesty EOS but with $\sim 5\,\%$
increased and decreased $Y_\mathrm{e}$ at nuclear density (sequences
$A1p$ and $A1m$, respectively). The details of this parametrization are
explained in Appendix~\ref{sec:yepar}, while the characteristics of
the models from these sequences are given in
Table~\ref{tab:injmodels}.

Fig.~\ref{fig:h_vs_t_A1O10.25} shows the GW strain versus time for
models $A1O10.25$, $A1O10.25s$, $A1O10.25m$, and $A1O10.25p$
during the late collapse, bounce, and early postbounce
phases. Although the behavior of the GW strain appears qualitatively
similar in these four cases, there are non-negligible quantitative
differences stemming from the changes in the EOS and $Y_e(\rho)$
parametrization, both of which grow with increasing postbounce time. 

The cyan pentagons in Fig.~\ref{fig:betaicbmeasuredMF} display the
measured $\beta_\mathrm{ic,b}$ as a function of the true
$\beta_\mathrm{ic,b}$ for sequence $A1s$. Despite the difference
between the two EOS, the matched filtering analysis measures
$\beta_\mathrm{ic,b}$ within $\lesssim 15\,\%$ of its correct value
for the waveforms. The average relative deviation between measured and
true $\beta_\mathrm{ic,b}$ for all $A1s$ models is $\sim 9\,\%$.  Such
small deviations are not surprising given the relatively weak
dependence of the GW signal features on the details of the EOS found
by \cite{dimmelmeier:08}.  The green stars and red triangles in
Fig.~\ref{fig:betaicbmeasuredMF} represent the measured
$\beta_\mathrm{ic,b}$ as a function of true $\beta_\mathrm{ic,b}$ for
sequences $A1m$ and $A1p$, respectively.  In the case of rapid
rotation ($\beta_\mathrm{ic,b} \gtrsim 0.05$), $\beta_\mathrm{ic,b}$
is extracted with $\lesssim 15\,\%$ accuracy with an average deviation
of $\sim 10\,\%$, only somewhat larger than in the case of ``known''
$Y_e$ parameterization.  Stochastic GW signal components from prompt
convection explain the outliers at small $\beta_\mathrm{ic,b}$.

Based on these results, we conclude that our matched filter analysis
can extract $\beta_\mathrm{ic,b}$ robustly with $\sim 20\,\%$ accuracy
for GW signals from rotating collapse, bounce, and ring-down
oscillations from galactic core-collapse events. This measurement is
rather robust and not very sensitive to uncertainties in inner-core
$Y_e$ and EOS.

\begin{figure}[t]
\centering
\includegraphics[width=0.97\columnwidth]{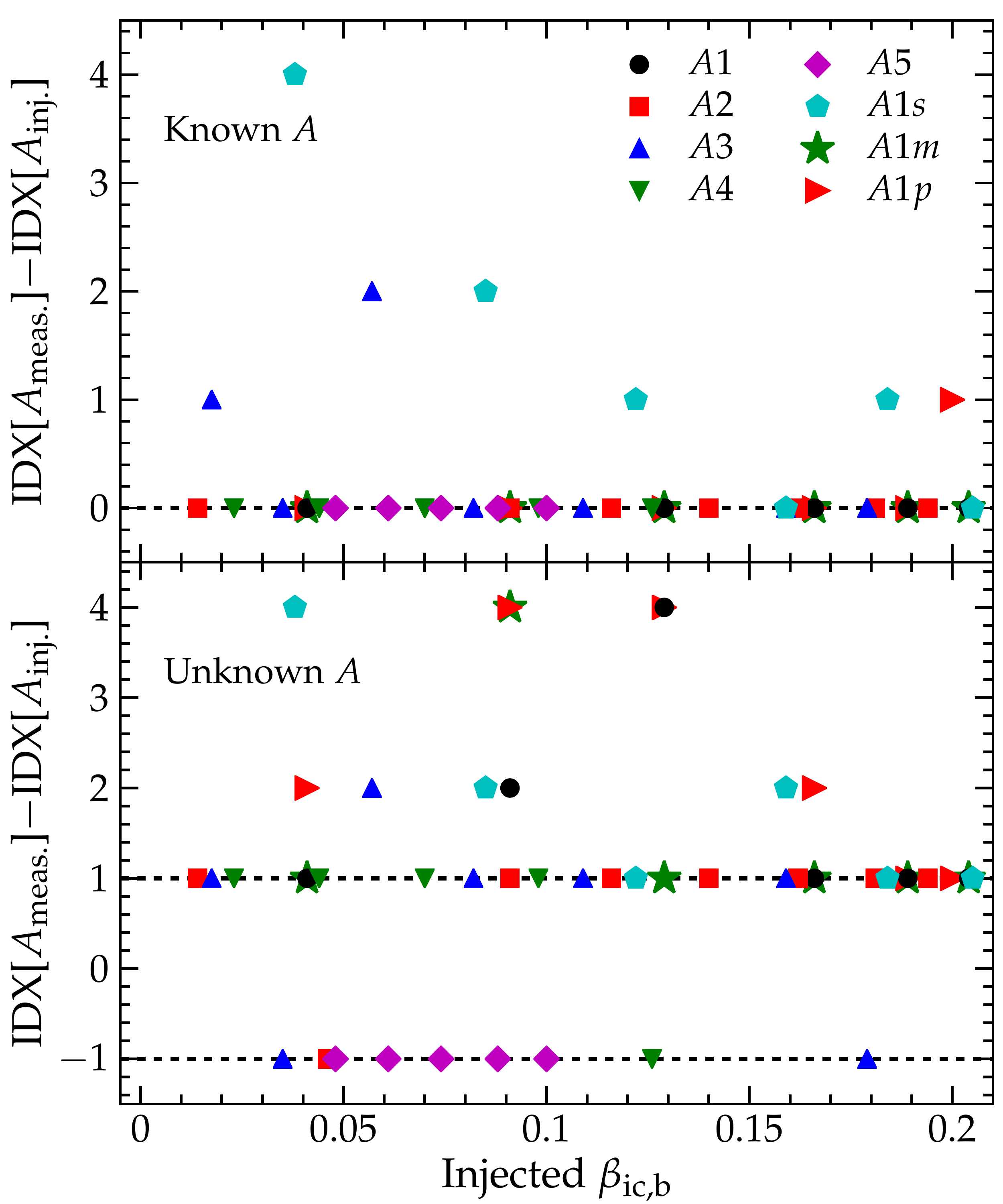}
\caption{$\delta i = \mathrm{IDX}[A_\mathrm{meas.}] -
  \mathrm{IDX}[A_\mathrm{inj.}]$ as a function of
  $\beta_\mathrm{ic,b}$. $\mathrm{IDX}[A_\mathrm{inj.}]$ and
  $\mathrm{IDX}[A_\mathrm{meas.}]$ denote the indices of
  the true and inferred values of differential rotation parameter $A$.
  As mentioned previously, $\delta i = 0$ ($\delta i \neq 0$) signify that $A$
  has been correctly (incorrectly) identified. The upper and lower panels represent 
  the cases in which the true value of $A$ is and isn't encompassed by the 
  template bank, respectively.}
\label{fig:AmeasuredMF}
\vspace{0.5ex}
\end{figure}

\subsubsection{Extraction of the Differential Rotation Parameter $A$}

The upper panel of Fig.~\ref{fig:AmeasuredMF} shows the quantity
$\delta i =
\mathrm{IDX}[A_\mathrm{meas.}]-\mathrm{IDX}[A_\mathrm{inj.}]$ as a
function of $\beta_\mathrm{ic,b}$, where
$\mathrm{IDX}[A_\mathrm{meas.}]$ is the integer index of the
differential rotation parameter $A_\mathrm{meas.}$ extracted by the
matched filter analysis.  $\mathrm{Idx.}[A_\mathrm{inj.}]$ is the
index of the true value of $A$ for the injected signal (e.g.,
$\mathrm{Idx.}[A] = 2$ for $A=A2$).  In this construction, $\delta i =
0$ ($\delta i \neq 0$) for the correct (incorrect) measurement of $A$.
It is important to point out a caveat in `measuring' the degree of
differential rotation using the method outlined here.  The
differential rotation law considered in this paper is somewhat
artificial, and it is not known if the cores of massive stars obey
this.  We therefore remind the reader that we present the ability to
measure the distribution of angular momentum in core-collapse
supernova progenitors, given that they obay the rotation law given by
Eq.~(\ref{eq:rotlaw}).

For sequences $A1$, $A4$, and $A5$, the values of $A$ are identified
correctly for all injected signals.  For sequences $A3$ ($A2$), $A$ is
determined accurately for $71\%$ ($88\%$) of injected waveforms.
Moreover, we find that $A$ corresponds to the next closest value in
all misidentification cases.  We note that misidentifications occur
only for slowly rotating models.  For $\beta_\mathrm{ic,b} \gtrsim
0.08$, $A$ is correctly determined for all injected waveforms.
  More slowly rotating models emit weaker GWs,
  so their signal-to-noise ratio in the detector is lower, which
  could be a potential cause of the misidentification. However, tests
  in which we placed such models at closer distances revealed that
  misidentifications occur even at high signal-to-noise ratio.  It is,
  hence, more likely that the convective component of the GW signal,
  which dominates in slowly rotating models, spoils the identification
  with the correct $A$. Our finding is also consistent with the notion
that the degree of differential rotation plays a significant role
only at rapid rotation (see Section \ref{sec:gw_peaks}).

For the $A1m$ and $A1p$ signals, in which $Y_e$ in the inner core is decreased and
increased, respectively, $A$ is inferred correctly for $100\%$ and $83\%$ of 
injections. We find that $A$ corresponds to the next closest value in the 
misidentification cases. 

For the $A1s$ signals, in which the
Shen~\emph{et~al.}~\cite{shen:98a,shen:98b} EOS is used in place of
the Lattimer \& Swesty EOS~\cite{lseos:91}, $A$ is inferred correctly
for only two rapidly rotating models ($\simeq 33\,\%$ of all models
from these sequences) with $\beta_\mathrm{ic,b}$ of $\sim 0.16$ and
$\sim 0.2$ (shown with cyan pentagons in Fig.~\ref{fig:AmeasuredMF}
and denoted as $A1s$). Another $\sim 33\,\%$ of models have $|\delta
i|=1$, while the remainder are measured with $|\delta i|>1$.
 This suggests that, unlike
  $\beta_\mathrm{ic,b}$, $A$ is rather sensitive to details of the
  nuclear EOS, and is thus more difficult to infer and features
  signifying different $A$ can be confused with features imprinted due
  to differences in the EOS. 
This is supported by the correct identification of $A$ only for rapid
rotation.  Rapidly rotating models reach lower maximum densities than
slowly spinning ones, where differences in EOS are less pronounced
than in the high density regime.  

To further test the robustness of our conclusions, we use GW
injections characterized by $A$ not represented in the template bank.
Here, we inject signals from model $Ai$, and filter the data only with
templates associated with models $Aj$, $j \neq i$.  The lower panel of
Fig.~\ref{fig:AmeasuredMF} presents $\delta i$ as a function of
$\beta_\mathrm{ic,b}$ for this scenario. Here, $\delta i = \pm 1$
implies $A_\mathrm{meas.}$ is estimated to be the closest available
value of $A$. For $A2$, $A4$ and $A5$, all injections are associated
with the closest $Ai$. For $A3$, $87\,\%$ of the injections are found
with $\delta i=\pm1$, while the remainder have $\delta i=2$. For
sequence $A1$, $67\%$ of injections return $\delta i = \pm1$, while
the remainder return $\delta i = 2,3$.  We find that for models $A1m$,
$A1p$, and $A1s$, $\delta i = \pm1$ for $67\,\%$, $33\,\%$ and
$50\,\%$ of injections, respectively.  Signals
  with EOS and $Y_\mathrm{e}(\rho)$ parametrization different from the
  fiducial ones result in a larger measurement error in the inferred
  value of $A$.

\subsection{Bayesian Model Selection}
\label{sec:bayesian}

We present now an alternative method to investigate the dependence 
of the features of the GW signal on the differential rotation parameter $A$ 
and its detectability.  We employ a Bayesian approach utilizing 
principal component analysis~\cite{mardia:80}, building upon previous 
work by R\"over~\emph{et~al.}~\cite{roever:09} and Logue
\emph{et~al.}~\cite{logue:12}. 

As discussed in previous sections, GW signals from progenitors
characterized by any given $A$ are expected to exhibit some strong
common features. To exploit this, we apply PCA to catalogs of
waveforms characterized by a common $A$ for each value of $A$. Principal
component analysis isolates dominant features of waveforms into linearly 
independent principal components, ordered by their
relevance. Mathematically, utilizing matrix $C$ containing a given
waveform catalog, one can factorize $C$ as 
\begin{equation}
C = U\Sigma V^{T}\,\,,
\end{equation}
where $U$ and $V$ are matrices comprised of the eigenvectors of
$CC^{T}$ and $C^{T}C$ respectively, and $\Sigma$ is a diagonal matrix,
composed from the square roots of corresponding eigenvalues. The principal
components, $U$, are organized according to their corresponding 
eigenvalues, such that the more dominant principal components (characterized 
by larger eigenvalues) are shifted to the first few columns of 
$U$.  Approximations to waveforms in $C$, in addition to arbitrary 
waveforms, can be constructed as
\begin{equation}
h_{i} \approx \sum_{j}U_{ij}\epsilon_{j}\,\,,
\end{equation}
where $h$ is the desired waveform approximation, and $\vec{\epsilon}$ contains
the projections of the original waveforms onto the $U$ basis, hereafter referred
to as principal component coefficients.

As in Section \ref{sec:templatebank}, we model the GW detector data,
$\tilde{d}$, as containing both some core-collapse supernova GW signal
$\tilde{h}(f;\vec{\lambda})$ and Gaussian noise $\tilde{n}$, colored
by known one-sided power spectral density (PSD) $S_{h}(f)$, where
$\tilde{d}$ and $\lambda$ are given by expressions (\ref{eq:di}) and
(\ref{eq:theta}) respectively.  We consider
  trial ``templates'' or signals, $\tilde{h}(f;\vec{\mu})$, where
\begin{align}
\vec{\mu} &= \{\vec{\epsilon},\vec{\lambda}\}\,\,,
\end{align}
which are reconstructed using principal components.

\begin{figure}[t]
\centering
\includegraphics[width=0.97\columnwidth]{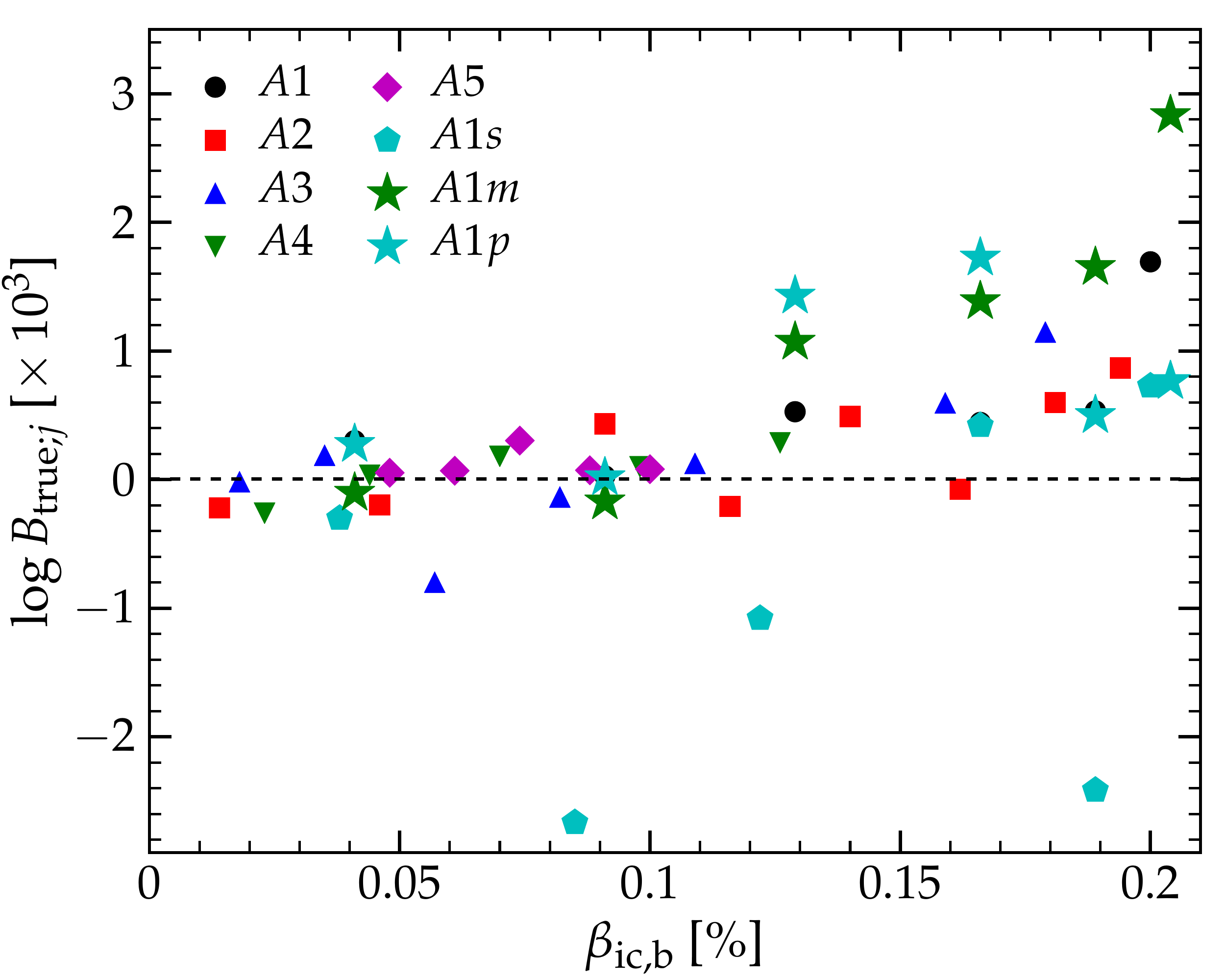}
\caption{$\log B_{\mathrm{true};j}$ for all injected waveforms. $\log
  B_{\mathrm{true};j} > 0$ and $\log B_{\mathrm{true};j} < 0$ imply
  correct and incorrect inference of $A$, respectively.  Large values
  of $\log B_{\mathrm{true};j}$ convey that the correct model has been
  chosen with a high degree of confidence.  The black dashed line
  denotes the confidence threshold $\log B_{\mathrm{true};j}=5$ (see
  main text for definition of this threshold). We see that $\log
  B_{\mathrm{true};j}$ increases with $\beta_\mathrm{ic,b}$, and at a
  given $\beta_\mathrm{ic,b}$, $A$ is inferred correctly with the
  highest confidence for injections associated with the strongest
  differential rotation ($A1$).  In the limit of slow rotation
  ($\beta_\mathrm{ic,b} \lesssim 0.05$), the correct model for $A$ is
  not determined for most injections.  Incorrect $A$
  is chosen for all injections simulated with the
  Shen~\emph{et~al.}~\cite{shen:98a,shen:98b} EOS ($A1s$), with the
  exception of two models characterized by extremely rapid rotation
  ($\beta_\mathrm{ic,b} \sim 0.16$).}
\label{fig:bayesfactor}
\vspace{0.5ex}
\end{figure}

Our goal is to compute the evidence, $p(d|\mathcal{H})$, that the data
observed contain a GW signal reconstructable from different sets of
principal components, each associated with a particular degree of
differential rotation.  The evidence, or marginal likelihood, of the
model $\mathcal{H}$ is calculated as
\begin{equation}
p(d|\mathcal{H}) = \int_{\vec{\mu}}
p(d|\vec{\mu};\mathcal{H})p(\vec{\mu}|\mathcal{H})\,d\vec{\mu}\,\,,
\end{equation}
where $p(\vec{\mu}|\mathcal{H})$ is the prior distribution on the parameters,
given the signal model (assumed to be flat in the absence of any physical
motivation to do otherwise) and $p(d|\vec{\mu};\mathcal{H})$ is the likelihood
function for the data.  Due to the Gaussian statistics of the noise, the
likelihood function for the presence of some signal 
$\tilde{h}(f_{i};\vec{\mu})$ can be written as
\begin{equation}
p(d|\vec{\mu};\mathcal{H}) =
\prod_{i}\,\frac{1}{\sigma_{i}\sqrt{2\pi}}\,\exp\left(-\frac{|\tilde{d}_{i} -
\tilde{h}(f_{i};\vec{\mu})|^{2}}{2\sigma_{i}^{2}}\right).
\end{equation}
Here, $\sigma_{i}^{2}$ is the variance of the noise in the $i^{\mathrm{th}}$
frequency bin, related to the PSD as 
\begin{equation}
S_{h}(f_{i}) = 2\frac{{\Delta t}^{2}}{T}\sigma_{i}^{2}\,\,,
\end{equation}
where $\Delta t$ and $T$ are the sampling time-step (the inverse of
the sampling frequency) and the total observation time, respectively.
To compute the evidence, we utilize an implementation of the nested
sampling algorithm~\cite{skilling:04}.  We perform an analysis closely
linked to previous work by~\cite{roever:09,logue:12}. We compute the
relative Bayes factor, $\log B_{i;j} = \log p(d|i) - \log p(d|j)$,
between models $i$ and $j$, to determine whether the evidence for
model $i$ is either greater than ($B_{i;j} > 0$) or less than
($B_{i;j} < 0$) the evidence for model $j$.  We compare a single
signal model $i$ to the noise model via $\log B_{i} = \log p(d|i) -
\log p(d,\mathrm{noise})$.

Connecting this to the physical motivation of our analysis, models $i$
and $j$ are PC sets constructed from waveforms catalogs characterized
by different values of $A$.  We `normalize' the Bayes factor for the
correct model $B_{\mathrm{true}}$ for each injected signal, to
illustrate whether the correct model for $A$ has been chosen. To do
this, we compute
\begin{equation}
\log B_{\mathrm{true};j} = \log B_{\mathrm{true}} - \max[\log B_{j}]\,\,, 
\end{equation}
where $\max[\log B_j]$ is the maximum logarithmic Bayes factor
obtained for values of $A$ other than the true one. $\log
B_{\mathrm{true};j} > 0$ ($\log B_{\mathrm{true};j} < 0$) states that
the correct model for $A$ has (has not) been inferred.  As common in
Bayesian model selection, we impose a confidence threshold $\eta$,
such that $\log B_{i;j} > \eta$ states that model $A_{i}$ is more
likely than $A_{j}$ with statistical significance. Following the
conventions in Logue \emph{et~al.}~\cite{logue:12}, we set $\eta
= 5$.

As in Section \ref{sec:templatebank}, we utilize simulated Gaussian
noise colored by the zero-detuned high power configuration of
aLIGO\,\cite{LIGO-sens-2010}, in the context of a single,
optimally-oriented GW detector.  We assume that the position,
inclination polarization of the source are known, such that the
antenna response functions are given by $F_{+} = 1$, $F_{\times} = 0$,
and place all sources at a known distance of $10\,\mathrm{kpc}$.
Limited by the size of the smallest waveform catalog, we use a subset
of $10$ PCs from each set to approximately 
  reconstruct injected waveforms.  Given this, the parameter space
  $\vec{\mu}$ is reduced to a $10$-dimensional subset, such that
  $\vec{\mu} \rightarrow \vec{\epsilon}$, where $\vec{\epsilon} =
  \{\epsilon_{1}, \ldots, \epsilon_{10}\}$.


\begin{longtable*}{lccccccccccccc}
  \caption{Summary of properties of models for
    injection. $\Omega_\mathrm{c}$ is the intitial central angular
    velocity, $\rho_\mathrm{c,b}$ is the central density at bounce,
    $\rho_\mathrm{c,pb}$ is the early postbounce central density,
    $\rho_\mathrm{max,pb}$ is the postbounce maximum density,
    $\beta_\mathrm{ic,b}$ and $\beta_\mathrm{ic,pb}$ ara ratios of the
    rotational kinetic energy to the gravitational binding energy of
    the inner core at bounce and early postbounce phase,
    respectively. $M_\mathrm{ic,b}$ and $J_\mathrm{ic,b}$ are the
    inner core mass and angular momentum at bounce, $|h_{+,2}|D$ is
    the second peak of the GW signal, while $|h_\mathrm{+,max}|D$ is
    its maximum value. $f_\mathrm{max}$ is the frequency at which the
    GW spectral energy density reaches a maximum
    value.} \label{tab:injmodels}\\ \hline\hline
  Model&$\Omega_\mathrm{c}$&$\rho_\mathrm{c,b}$&$\rho_\mathrm{c,pb}$&$\rho_\mathrm{max,pb}$&$\beta_\mathrm{ic,b}$&$\beta_\mathrm{ic,pb}$&$M_\mathrm{ic,b}$&$J_\mathrm{ic,b}$&${H}D$&$|h_\mathrm{+,max}|D$&$E_\mathrm{GW}$&$f_\mathrm{max}$\\ &[$\mathrm{rad\,s^{-1}}$]&[$10^{14}$
    &[$10^{14}$ &[$10^{14}$
         &[$10^{-2}$]&[$10^{-2}$]&[$M_\odot$]&[$10^{48}$
           &[cm]&[cm]&[$10^{-9}$ &[Hz]\\ &
             &$\mathrm{gcm^{-3}}$]&$\mathrm{g\,cm^{-3}}$]&$\mathrm{g\,cm^{-3}}$]&
       & & &$\mathrm{erg\,s}$]& & &$M_\odot c^2$]& \\
  \hline
  \endfirsthead
  \caption{Continued}\\
  \hline\hline
  Model&$\Omega_\mathrm{c}$&$\rho_\mathrm{c,b}$&$\rho_\mathrm{c,pb}$&$\rho_\mathrm{max,pb}$&$\beta_\mathrm{ic,b}$&$\beta_\mathrm{ic,pb}$&$M_\mathrm{ic,b}$&$J_\mathrm{ic,b}$&${H}D$&$|h_\mathrm{+,max}|D$&$E_\mathrm{GW}$&$f_\mathrm{max}$\\
       &[$\mathrm{rad\,s^{-1}}$]&[$10^{14}$          &[$10^{14}$            &[$10^{14}$            &[$10^{-2}$]&[$10^{-2}$]&[$M_\odot$]&[$10^{48}$        &[cm]&[cm]&[$10^{-9}$    &[Hz]\\
       &                        &$\mathrm{gcm^{-3}}$]&$\mathrm{g\,cm^{-3}}$]&$\mathrm{g\,cm^{-3}}$]&           &           &           &$\mathrm{erg\,s}$]&    &    &$M_\odot c^2$]&    \\
  \hline
  \endhead
  \hline
  \hline
  \multicolumn{11}{r}{Continued on Next page...}
  \endfoot
  \hline
  \hline
  \endlastfoot
$A1O5.25  $&  5.25  & 4.28 & 3.37 & 3.37 &  4.09  &  3.36  & 0.63 & 1.83 & 111.19\z & 189.59 & 16.19 & 991.53 \\
$A1O5.25m $&  5.25  & 4.18 & 3.41 & 3.41 &  4.04  &  3.38  & 0.60 & 1.68 & 93.98    & 163.91 & 14.33 & 864.48 \\
$A1O5.25p $&  5.25  & 4.18 & 3.30 & 3.30 &  4.14  &  3.33  & 0.60 & 1.96 & 132.49\z & 213.63 & 23.23 & 950.30 \\
$A1O5.25s $&  5.25  & 3.33 & 2.65 & 2.65 &  3.79  &  3.15  & 0.60 & 1.72 & 87.97    & 158.17 & 11.41 & 687.21 \\
$A1O8.25  $&  8.25  & 3.87 & 3.00 & 3.03 &  9.12  &  7.33  & 0.68 & 3.45 & 252.70\z & 436.55 & 70.16 & 826.04 \\
$A1O8.25m $&  8.25  & 3.88 & 3.05 & 3.08 &  9.02  &  7.37  & 0.68 & 3.21 & 191.78\z & 377.27 & 52.56 & 847.37 \\
$A1O8.25p $&  8.25  & 3.88 & 2.94 & 2.96 &  9.19  &  7.29  & 0.68 & 3.58 & 302.69\z & 469.06 & 86.19 & 783.88 \\
$A1O8.25s $&  8.25  & 2.98 & 2.45 & 2.45 &  8.46  &  6.87  & 0.66 & 3.10 & 218.82\z & 368.80 & 49.02 & 737.13 \\
$A1O10.25 $& 10.25\z& 3.42 & 2.65 & 2.72 & 12.90\z& 10.20\z& 0.76 & 4.81 & 199.43\z & 521.24 & 87.74 & 645.55 \\
$A1O10.25m$& 10.25\z& 3.45 & 2.75 & 2.80 & 12.80\z& 10.20\z& 0.74 & 4.59 & 180.30\z & 490.37 & 74.00 & 922.20 \\
$A1O10.25p$& 10.25\z& 3.45 & 2.59 & 2.65 & 13.00\z& 10.10\z& 0.74 & 4.92 & 228.11\z & 540.91 & 93.17 & 541.00 \\
$A1O10.25s$& 10.25\z& 2.76 & 2.22 & 2.26 & 12.20\z&  9.64  & 0.72 & 4.45 & 211.99\z & 492.01 & 73.49 & 795.74 \\
$A1O12.25 $& 12.25\z& 3.06 & 2.44 & 2.47 & 16.60\z& 13.10\z& 0.81 & 6.37 & 202.70\z & 541.45 & 73.49 & 475.36 \\
$A1O12.25m$& 12.25\z& 3.06 & 2.49 & 2.53 & 16.50\z& 13.20\z& 0.80 & 6.22 & 177.02\z & 559.48 & 73.84 & 492.63 \\
$A1O12.25p$& 12.25\z& 3.06 & 2.33 & 2.36 & 16.40\z& 13.00\z& 0.80 & 6.47 & 174.29\z & 512.50 & 55.90 & 432.05 \\
$A1O12.25s$& 12.25\z& 2.51 & 1.97 & 2.05 & 15.90\z& 12.50\z& 0.77 & 5.80 &-171.29\pz& 522.33 & 66.92 & 486.68 \\
$A1O13.75 $& 13.75\z& 2.83 & 2.22 & 2.25 & 18.90\z& 15.40\z& 0.83 & 7.28 & 168.01\z & 492.28 & 44.51 & 430.02 \\
$A1O13.75m$& 13.75\z& 2.88 & 2.32 & 2.35 & 19.10\z& 15.50\z& 0.85 & 7.72 & 182.76\z & 547.19 & 56.37 & 440.95 \\
$A1O13.75p$& 13.75\z& 2.88 & 1.94 & 1.97 & 18.30\z& 14.40\z& 0.85 & 7.63 & 129.49\z & 398.03 & 19.80 & 581.74 \\
$A1O13.75s$& 13.75\z& 2.36 & 1.89 & 1.92 & 18.40\z& 14.70\z& 0.81 & 7.03 & 158.99\z & 511.13 & 53.99 & 435.78 \\
$A1O15.25 $& 15.25\z& 2.49 & 1.79 & 1.82 & 20.40\z& 16.50\z& 0.88 & 9.00 &  94.80   & 369.07 & 13.96 & 275.55 \\
$A1O15.25m$& 15.25\z& 2.61 & 2.06 & 2.10 & 21.10\z& 17.60\z& 0.89 & 9.19 & 160.36\z & 469.33 & 31.17 & 382.93 \\
$A1O15.25p$& 15.25\z& 1.14 & 1.16 & 2.10 & 19.40\z& 14.50\z& 0.89 & 9.09 & -59.28\p & 262.53 &\z6.77 & 221.11 \\
$A1O15.25s$& 15.25\z& 2.28 & 1.65 & 1.70 & 20.50\z& 16.40\z& 0.85 & 8.40 & 123.21\z & 469.88 & 25.96 & 353.33 \\
\end{longtable*}

\begin{figure}[t]
\centering
\includegraphics[width=0.97\columnwidth]{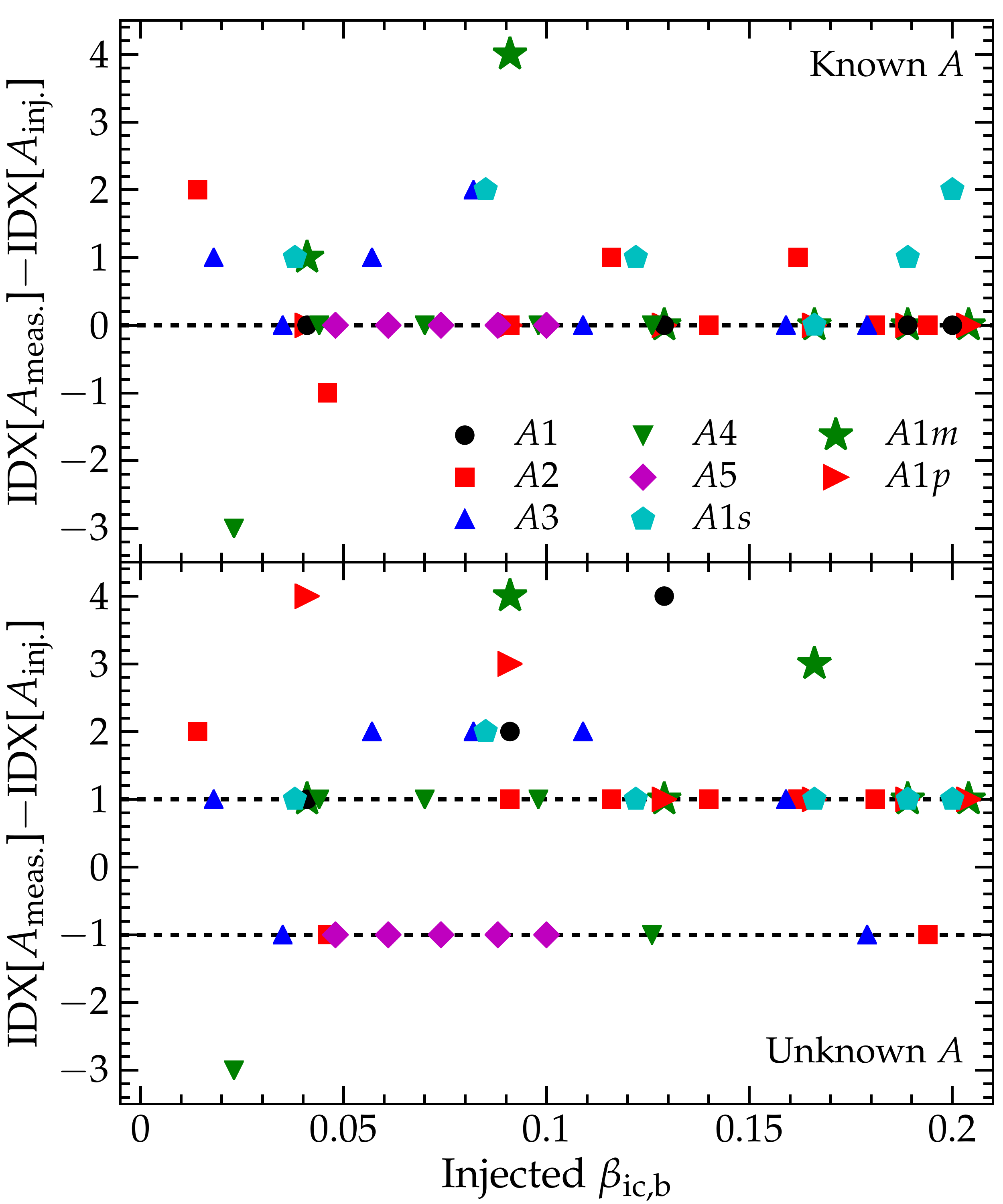}
\caption{The quantity $\delta i =
  \mathrm{Idx.}[A_\mathrm{meas.}]-\mathrm{Idx.}[A]$ as a function of
  $\beta_\mathrm{ic,b}$, where $\mathrm{Idx.}[A_\mathrm{meas.}]$ is
  the index of the differential rotation parameter $A_\mathrm{meas.}$
  inferred by the model selection analysis, and $\mathrm{Idx.}[A]$ is the
  index of the true value of $A$ (e.g., $\mathrm{Idx.}[A] = 2$ for $A=A2$). 
  The upper and lower panels display results for cases in which $A$ for all
  injections is known and unknown, respectively.}
\label{fig:AmeasuredBF}
\vspace{0.5ex}
\end{figure}

We construct PCs using the model waveforms described in
Tables~\ref{tab:models} and \ref{tab:results}. The injected signals
are the same used for injection in Section~\ref{sec:templatebank},
which are distinct from those used to generate the PCs.

Figure~\ref{fig:bayesfactor} presents the normalized $\log
  B_{\mathrm{true},j}$ for all injected waveforms.
Large values of $\log B_{\mathrm{true},j}$ indicate a high degree of
confidence in the chosen model for $A$.  The dashed black line
represents the detectability threshold $\log B_{\mathrm{true},j}=5$
discussed above. At $\beta_\mathrm{ic,b} \lesssim 0.05 $, most
injected signals across all $A$ have negative $\log
B_{\mathrm{true},j}$, suggesting that it is difficult to infer the
correct model for $A$ in the slow rotation
limit. Test with sources located at closer
  distances show that this is not a consequence of the low
  signal-to-noise ratio of the GW signal emitted by such models.  Instead, it
  is most likely due to a combination of the facts that (\emph{i}) the
  stochastic GW signal from prompt convection is comparable to or
  stronger than the signal from collapse, bounce, and ring-down, and,
  (\emph{ii}) that there is little influence of $A$ on the magnitude
  of the peaks of the GW signal at slow rotation (see
  Section~\ref{sec:gw_peaks}).

For a given $\beta_\mathrm{ic,b}$, model $A1$, which is the most
strongly differentially rotating, has the largest $\log
B_{\mathrm{true},j}$, suggesting that the ability to infer $A$ with
this method is greatest in extremely differentially rotating models.
We also see that the magnitude of $\log B_{\mathrm{true},j}$ tends to
grow with increasing $\beta_\mathrm{ic,b}$, and the correct model for
$A$ is determined for the majority of injections with 
$\beta_\mathrm{ic,b} \gtrsim
0.08$. This is in agreement with our GW peaks analysis in
Section~\ref{sec:gw_peaks}, where significant dependence on $A$ is
observed in the large $\beta_\mathrm{ic,b}$ regime. 

To test the robustness of this conclusion, we inject the waveform set
$A1s$ simulated with the Shen~\emph{et~al.}
EOS~\cite{shen:98a,shen:98b}, described in
Section~\ref{sec:templatebank}.  We find that in this case, the
correct model for $A$ is determined only for two models with very
rapid rotation (with $\beta_\mathrm{ic,b}$ of $\sim 0.16$ and $\sim
0.2$). The maximum densities reached in these cases are relatively
low, and the two EOS are not very different in this regime. In the
slow rotation limit, the injected signals are strongly associated with
incorrect models for $A$.  
This suggests that, if the differences between the true nuclear EOS
and that used for PC construction are of the same order as the
differences between the Lattimer \& Swesty~\cite{lseos:91} and
Shen~\emph{et~al.}~\cite{shen:98a,shen:98b} EOS, then the inference of
the progenitor's angular momentum distribution from the GW signal
observed is significantly more difficult than if the nuclear EOS was
known.  This conclusion is consistent with that of the matched filter
analysis presented in the previous Section~\ref{sec:templatebank}.

We also inject waveform sets $A1m$ and $A1p$, simulated with modified
$Y_\mathrm{e}(\rho)$ parametrization, as explained in
Section~\ref{sec:templatebank}.  For these injections (marked with
large green stars and red triangles, respectively, in
Fig.~\ref{fig:bayesfactor}), the correct model for $A$ is determined
in the limit of fast rotation, whereas the wrong model is chosen in
the slow rotation regime.  In addition, we find that for models with
correctly chosen $A$, the magnitude of $\log B_{i,j}$ is smaller than
for injections from the $A1$ sequence where the $Y_\mathrm{e}(\rho)$
parametrization is ``known''.  This suggests that unless
$Y_{\mathrm{e}}$ in the inner core is known to within $5\%$ accuracy,
our ability to infer the correct model for $A$ suffers greatly.

The upper panel of Fig.~\ref{fig:AmeasuredBF} shows $\delta i =
\mathrm{IDX}[A_\mathrm{meas.}]-\mathrm{IDX}[A_\mathrm{inj.}]$ (as
previously defined) as a function of $\beta_\mathrm{ic,b}$ for the
injected waveforms.  Here, $\mathrm{IDX}[A_\mathrm{meas.}]$ is the
index of the differential rotation parameter $A_\mathrm{meas.}$
determined by Bayesian model selection, where $\log
B_{\mathrm{meas,j}} > 5$ by definition.  For most injections, $\delta
i = 0$, signifying that the correct model for $A$ has been inferred.
We note that the fraction of injections with $\delta i = 0$ grows with
increasing $\beta_\mathrm{ic,b}$, which is consistent with the
previous conclusion that the ability to determine $A$  is
greater for rapidly rotating models. For models $A1$, $A2$, $A3$,
$A4$, $A5$, $A$ is measured correctly in $\simeq 100\,\%$, $\simeq
50\,\%$, $\simeq 57\,\%$, $\simeq 80\,\%$, $\simeq 100\,\%$ of
cases. For sequences $A1m$, $A1p$, and $A1s$, $A$ is correctly
inferred in $\simeq 67\,\%$, $\simeq 100\,\%$ and $\simeq 17\,\%$ of
cases, respectively.  These values are consistent with those obtained
from the matched filter analysis.

The lower panel of Fig.~\ref{fig:AmeasuredBF} shows $\delta i$ as a
function of $\beta_\mathrm{ic,b}$ for the case of injections with
``unknown'' $A$, for which the correct model for $A$ is excluded from
the model selection analysis.  In this case, we see that the majority
of injected models have $\delta i = \pm 1$, implying identification
with the closest $A$ to that injected. Measurement of $A$ once more
improves with incresing $\beta_\mathrm{ic,b}$.

To conclude this section, we note that a
  directly comparable analysis for $\beta_{\mathrm{ic,b}}$ is not
  possible, since for a given $A$, many parameters affect
  $\beta_{\mathrm{ic,b}}$, such as $\Omega_{\mathrm{c}}$, EOS, and
  $Y_\mathrm{e}(\rho)$ parametrization.  However, a roughly analogous
  analysis could be constructed in which ``models'' describe ranges of
  $\beta_{\mathrm{ic,b}}$ (e.g., $0 \leq \beta_{\mathrm{ic,b}} \leq
  0.05$) rather than discrete values.  This blurs the line
  between model selection and parameter estimation, since the proposed
  ``models'' are just subsets of one model for $\beta_{\mathrm{ic,b}}$, 
  rather than different models. Instead, we try to estimate some range on 
  $\beta_{\mathrm{ic,b}}$, within which it is most likely to be.
  Well-posed Bayesian methods for
  parameter estimation typically require a continuous analytical model
  describing the parameter dependence of the
  system~\cite{skilling:04}. Such an analytic model (which could, e.g.,
  be constructed by interpolating between discrete waveforms) is not
  presently available to us. Hence, we choose to postpone Bayesian
  parameter estimation of $\beta_\mathrm{ic,b}$ to future work.


\section{Summary and Conclusions}
\label{sec:summary}

Observations of stellar surface velocities show that most massive
stars rotate and some do so with velocities close to break-up (e.g.,
\cite{ramirezagudelo:13,huang:10}). The internal distribution of
angular momentum is, however, rather uncertain and this is true in
particular for the cores of presupernova stars. Rotation can influence
the collapse, bounce, and postbounce dynamics and may play a role in
driving the explosion. It is thus important to understand, or better,
measure the angular momentum distribution in the core of massive
stars. As we have shown in this paper, the observation of
gravitational waves (GWs) from the next galactic core-collapse
supernova may offer us the opportunity to do just that.

We have carried out an extensive set of axisymmetric
general-relativistic simulations of rotating core collapse to study
the influence of the angular momentum distribution on the GW signal of
rotating collapse, bounce, and the very early postbounce ring-down
phase. In total, we have simulated 124 different models,
systematically probing the effects of ``total rotation''
(parameterized either by the angular momentum of the homologous inner
core at bounce or by $\beta_\mathrm{ic,b} = T/|W|\big|_\mathrm{ic,b}$)
and the precollapse degree of differential rotation. We have also
performed simulations with a different nuclear equation of state
(EOS), variations in the electron fraction of the inner core, and
increased numerical resolution to test for systematic
uncertainties. We have employed a single presupernova stellar model,
since we have previously (in \cite{ott:12a}) shown that for a given
angular momentum distribution as a function of enclosed mass, EOS, and
electron-capture treatment, the universal nature of core collapse
\cite{goldreich:80,yahil:83} washes out variations due to differences
in precollapse progenitor structure.

Our results show that the overall dynamics of rotating core collapse
is rather insensitive to the precise distribution of angular momentum
within the inner core. We find that there is a simple linear mapping
between the two total rotation measures $J_\mathrm{ic,b}$ and
$\beta_\mathrm{ic,b}$ and the centrifugally-enhanced mass of the inner
core at bounce ($M_\mathrm{ic,b}$) throughout most of the explored
parameter space. Variations in the angular momentum distribution
become relevant to the detailed dynamics of collapse and bounce only in
very rapidly rotating cases with $\beta_\mathrm{ic,b} \gtrsim 0.13 -
0.15$, which corresponds to an inner core angular momentum at bounce
of $J_\mathrm{ic,b} \gtrsim 5-6\times
10^{48}\,\mathrm{erg}\cdot\mathrm{s}$ and early postbounce
density-weighted average core spin periods of $\lesssim
8-10\,\mathrm{ms}$.  While unimportant for the overall dynamics,
differential rotation does affect the structure and postbounce
evolution of the protoneutron star even in more slowly spinning
cores. At fixed total rotation at bounce, more differentially rotating
inner cores have more centrifugally-deformed (oblate) innermost
regions while their overall shape is less oblate than that of their
more uniformly spinning counterparts that have more centrifugal
support at greater radii (and enclosed-mass coordinates).

In slowly rotating models ($\beta_\mathrm{ic,b} \lesssim 0.05$), the
degree of precollapse differential rotation has little influence on
the GW signal and there are simple linear relationships that allow one
to map back from the amplitude of the pronounced and easily
identifiable bounce peak $h_\mathrm{1,neg}$ to $J_\mathrm{ic,b}$ and
$\beta_\mathrm{ic,b}$: $J_\mathrm{ic,b} \approx 10^{48}
(h_{1,\mathrm{neg}} D /
100\,\mathrm{cm})\,\mathrm{erg}\cdot\mathrm{s}$ and
$\beta_\mathrm{ic,b} \approx 2.3\times 10^{-2} (h_{1,\mathrm{neg}} D /
100\,\mathrm{cm})$. For this purpuse, the distance $D$ must be known
with good accuracy, which is likely for the next galactic
core-collapse supernova.

The structural changes due to differential rotation have important
ramifications for the GW signal in more rapidly spinning models with
$\beta_\mathrm{ic,b} \gtrsim 0.05-0.08$ ($J_\mathrm{ic,b} \gtrsim
2-3\times 10^{48}\,\mathrm{erg}\cdot\mathrm{s}$), corresponding to
early-postbounce protoneutron star spin periods of $\lesssim
12-16\,\mathrm{ms}$. More differentially rotating models yield higher
global peak GW strain amplitudes at bounce and emit more energy in
GWs. Total rotation \emph{and} the degree of differential rotation
influence the values of the first three local extrema of the GW
signal, $h_\mathrm{1,pos}$, $h_\mathrm{1,neg}$, $h_\mathrm{2,pos}$, in
a highly systematic way.

\begin{figure}[t]
\centering
\includegraphics[width=0.97\columnwidth]{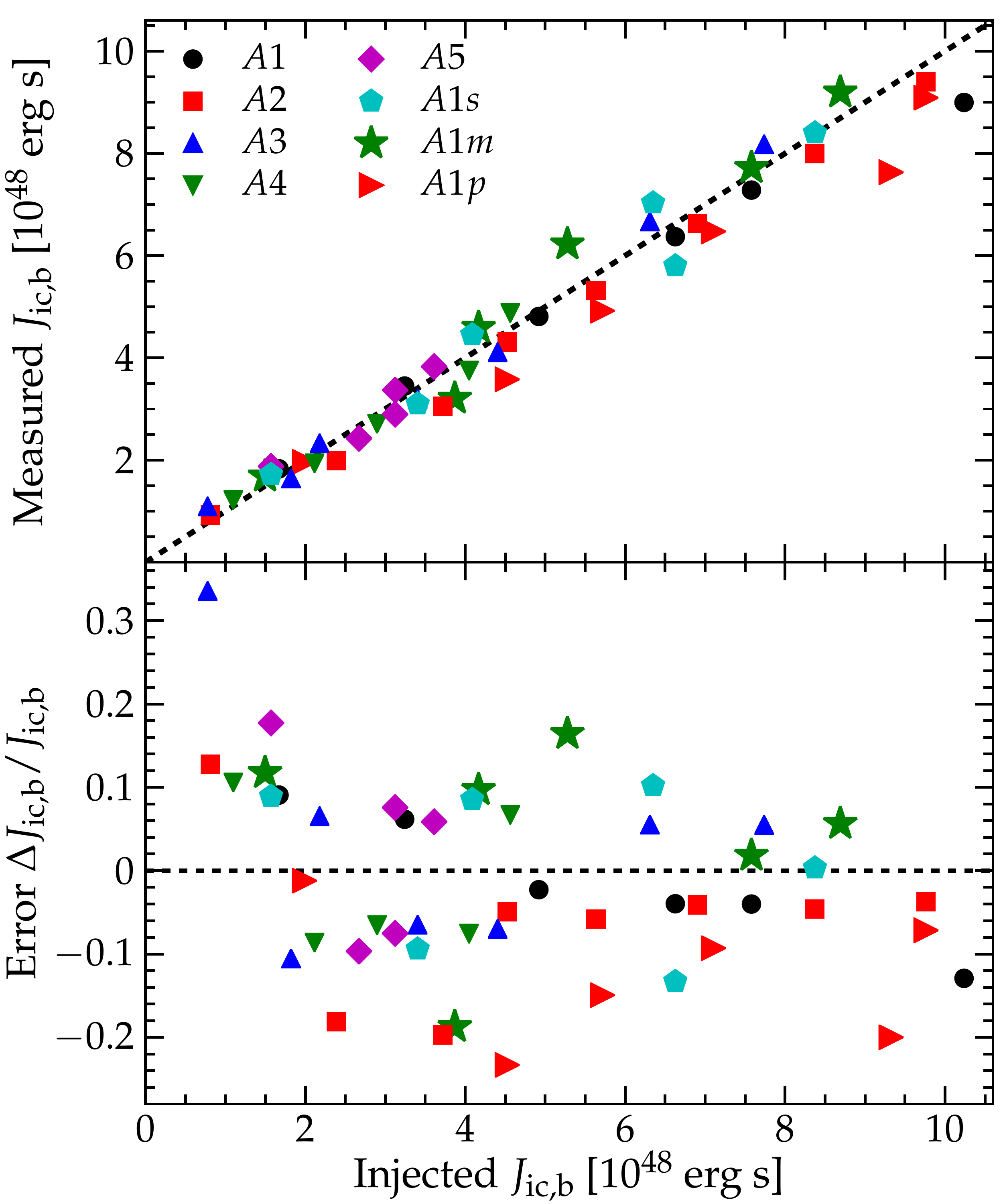}
\caption{Results of our matched filtering analysis
  (Section~\ref{sec:templatebank}) for the angular momentum of the
  inner core at bounce ($J_\mathrm{ic,b}$). Top panel: Extracted
  $J_\mathrm{ic,b}$ as a function of $J_\mathrm{ic,b}$ corresponding
  to the injected waveform. Bottom panel: relative measurement
  error. This analysis assumes optimal source-detector orientation and
  a source distance of $10\,\mathrm{kpc}$. The different symbols
  correspond to models with different degree of differential rotation
  as given by the legend. The $A1s$ models are $A1$ models, but
  evolved with the Shen EOS~\cite{shen:98b,hshen:11} and the $A1m$ ($A1p$) models used
  a $Y_e(\rho)$ parameterization during collapse that was reduced
  (increased) by $5\%$ near nuclear density compared to the fiducial
  one. Our results show that -- in the optimal case considered here --
  one can measure the angular momentum of the inner core at bounce
  with $\sim$20-30\% accuracy for a rapidly spinning galactic
  core-collapse supernova.}
\label{fig:Jmeasured}
\vspace{0.5ex}
\end{figure}

We have exploited this systematic dependence. Our results show that it
is possible to extract both total rotation (both $\beta_\mathrm{ic,b}$
and $J_\mathrm{ic,b}$, since the two are simply related) and the
degree of differential rotation from a previously \emph{unknown}
observed galactic rotating core collapse GW signal from a source at a
\emph{known} distance of $D=10\,\mathrm{kpc}$ via simple
cross-correlation with waveforms from a numerical template GW signal
bank created from our model GW signals. Since more rapidly spinning
cores have a smaller contribution to their GW signals from stochastic
convective motions, this works best for rapid rotation and our matched
filtering analysis can \emph{measure} total rotation to within
$\sim20\%$ for a rapidly rotating ($\beta_\mathrm{ic,b} \gtrsim 0.08$,
$J_\mathrm{ic,b} \gtrsim 3\times
10^{48}\,\mathrm{erg}\cdot\mathrm{s}$) core at $D=10\,\mathrm{kpc}$
that is optimally oriented with respect to a single GW
detector. Measuring total rotation is also possible for more slowly
spinning cores, though the errors may be $\gtrsim
25-35\%$. Figure~\ref{fig:Jmeasured} shows the $J_\mathrm{ic,b}$
inferred by our matched-filtering analysis as a function of the true
$J_\mathrm{ic,b}$ associated with each injected waveform. The injected
waveforms are not part of the template bank used. Thus, this
represents the realistic case that the exact waveform is not known.

For rapidly rotating cores ($\beta_\mathrm{ic,b}\gtrsim 0.08$) the
differential rotation parameter $A$ of the employed rotation law can
be extracted with good precision (maximum offset of $A{i}$ in $i$ is
$\pm$1).  We find the same result if we instead apply principal
component analysis and Bayesian model selection for the five choices
of differential rotation parameter $A{i}, i \in [1,5]$ that we
consider in this study.

While our simulations are numerically well converged, our tests reveal
important systematic uncertainties associated with the nuclear EOS and
the electron fraction $Y_e$ in the inner core at bounce. We find that
a $\pm 5\%$ variation of $Y_e$ or a change of the EOS from
Lattimer-Swesty with $K=220\,\mathrm{MeV}$ \cite{lseos:91} to the
H.~Shen EOS \cite{shen:98a,hshen:11} can spoil the accuracy with which
we can extract total and differential rotation.

The EOS dependence of our results underlines the need for improved
nuclear EOS tables that take into account all new experimental,
observational, and theoretical EOS constraints
\cite{lattimer:13}. Future simulations of rotating core collapse
should also consider a broader range of nuclear EOS models (e.g.,
\cite{gshen:11a,gshen:11b,steiner:13b}) to further explore the
sensitivity of the GW signal to the nuclear EOS.

Addressing uncertainties in the $Y_e$ of the inner core will
ultimately require full neutrino radiation-hydrodynamics simulations
with up-to-date electron capture rates for heavy nuclei (e.g.,
\cite{juodagalvis:10}), full velocity dependence, and inelastic
neutrino-electron scattering, which all have an effect on the $Y_e$ in
the inner core (e.g., \cite{lentz:12a}). Such simulations, while
computationally extremely intense, are possible now, for example with
the radiation-hydrodynamic variant of the {\tt CoCoNuT} code developed
by B.~M\"uller~\cite{mueller:09phd}.

\vskip.2cm 

In this study, we have broken entirely new ground by combining
precision computational waveform modeling with methods of GW
astronomy. We have given the proof of principle that information on
both total and differential rotation can be extracted, or at least
constrained, from the GW signal of the next galactic core collapse
event. Future work must address our study's many deficiencies. The
most important of these may be: (\emph{i}) Although axisymmetry is an
excellent approximation for collapse, bounce, and rign-down
oscillations for rotating axisymmetric progenitors, the subsequent
postbounce evolution (not considered in this work) is likely to
exhibit nonaxisymmetric features in the GW
signal~\cite{kuroda:13,ott:07prl,scheidegger:10b}. Moreover,
Kuroda~\emph{et al.} have argued that nonaxisymmetric perturbations in
the inner core may alter the bounce and postbounce gravitational wave
signal. It is presently unclear if such perturbations are present in
the core, but they are likely to be present in the shell burning
layers surrounding the core \cite{arnett:11a,couch:13d}, which are
irrelevant for the GW signals studied here.  (\emph{ii}) We considered
only a single rotation law, but realistic cores of massive stars do
not necessarily follow it. (\emph{iii}) We assumed optimal
source-detector alignment and only a single detector with Gaussian
noise. A real core collapse event is unlikely to be optimally aligned,
but a network of second-generation detectors can mitigate reduced
signal strength due to misalignment. (\emph{iv}) We assumed the
distance to the source to be known precisely. For a real core collapse
event, the distance is unlikely to be known exactly.  (\emph{v}) Our
treatment of electron capture during collapse relies on a
single-parameter density fit of $Y_e(\rho)$ from spherically symmetric
radiation-hydrodynamics simulations. Rapid rotation may lead to
significant deviations from such simple fits in full 2D
radiation-hydrodynamics simulations and this could have a significant
quantitative effect on the predicted GW signals.

\vspace*{0.5cm}
{\begin{center} \bf ACKNOWLEDGMENTS \end{center}}

We thank P.~Cerda-Duran, S.~Couch, J.~Clark, S.~de~Mink, B.~Engels,
R.~Haas, I.~S.~Heng, H.~Klion, J.~Logue, P.~M\"osta, B.~M\"uller,
J.~Novak, E.~O'Connor, U.~C.~T.~Gamma, C.~Reisswig, L.~Roberts, and
A.~Weinstein for helpful discussions.  This work is supported by the
National Science Foundation under grant numbers PHY-1151197,
AST-1212170, OCI-0905046, and PHY-1068881, by the Sherman Fairchild
Foundation, and the Alfred~P.~Sloan Foundation. Results presented in
this article were obtained through computations on the Caltech
computer cluster ``Zwicky'' (NSF MRI award No.\ PHY-0960291), on the
NSF XSEDE network under grant TG-PHY100033, on machines of the
Louisiana Optical Network Initiative, and at the National Energy
Research Scientific Computing Center (NERSC), which is supported by
the Office of Science of the US Department of Energy under contract
DE-AC02-05CH11231.


\begin{appendix}

\section{$Y_e(\rho)$ parametrized deleptonization scheme}
\label{sec:yepar}

Following~\cite{liebendoerfer:05fakenu}, we use the following fitting
function to model the functional dependence of $Y_\mathrm{e}$ on $\rho$: 
\begin{eqnarray}
Y_\mathrm{e} &=& \frac{1}{2} \left( Y_\mathrm{e,1} + Y_\mathrm{e,1}
\right) + \frac{x}{2} \left( Y_\mathrm{e,1} -
Y_\mathrm{e,1} \right) \\\nonumber &+& Y_\mathrm{e,c}
\left[1-|x|+4|x|(|x|-1/2)(|x|-1)\right], 
\end{eqnarray}
where
\begin{eqnarray}
x = \mathrm{max} \left[-1, \mathrm{min} \left(1, \frac{2\log\rho -
    \log\rho_2 - \log\rho_1} {\log\rho_2-\log\rho_1} \right)\right] 
\end{eqnarray}
and $\rho_1 = 10^7\,\mathrm{g\,cm^{-3}}$, $\rho_2 =
10^{13}\,\mathrm{g\,cm^{-3}}$, $Y_\mathrm{e,1}=0.5$,
$Y_\mathrm{e,2}=0.29$, and $Y_\mathrm{e,c}=0.035$. When density
$\rho$ is above $\rho_2$, we make the following correction to
$Y_\mathrm{e}$:
\begin{eqnarray}
Y_\mathrm{e} = Y_\mathrm{e}(\rho_2) 
+\frac{\log\rho - \log\rho_2}
     {\log\rho_\mathrm{cor}-\log\rho_2}
     \left[Y_\mathrm{e,cor}-Y_\mathrm{e}(\rho_2)\right],
\end{eqnarray}
where $Y_\mathrm{e,cor}$ is chosen to be $0.2717$ for our fiducial
$Y_\mathrm{e}(\rho)$ parametrization. In our $5\,\%$ reduced
(increased) $Y_\mathrm{e}(\rho)$ parametrization, we use a $5\,\%$
smaller (larger) value of $Y_\mathrm{e,cor}$.   

\end{appendix}


\end{document}